\documentclass[useAMS,usenatbib]{mn2e}

\usepackage{amssymb}
\usepackage{graphicx}
\newcommand{\src}{Swift~J1753.5--0127}

\title[Disc-jet coupling in black hole candidates]{Investigating the disc-jet coupling in accreting compact objects using the black hole candidate Swift~J1753.5-0127}
\author[P. Soleri et al.]{P. Soleri$^{1,2}$\thanks{E-mail:
soleri@astro.rug.nl}, R. Fender$^{3,1}$, V. Tudose$^{4,5,6}$, D. Maitra$^{7,1}$, M. Bell$^{3}$, M. Linares$^{8,1}$, \newauthor
D. Altamirano$^{1}$, R. Wijnands$^{1}$, T. Belloni$^{9}$, P. Casella$^{3,1}$, J.~C.~A. Miller-Jones$^{10}$, \newauthor
T. Muxlow$^{11}$, M. Klein-Wolt$^{12,1}$, M. Garrett$^{4,13,14}$ and M. van der Klis$^{1}$\\
$^{1}$Astronomical Institute Anton Pannekoek, University of Amsterdam, Science Park 904, 1098 XH, Amsterdam, The Netherlands\\
$^{2}$Kapteyn Astronomical Institute, University of Groningen, PO Box 800, 9700 AV, Groningen, The Netherlands\\
$^{3}$School of Physics and Astronomy, University of Southampton, Hampshire, SO17 1BJ, UK\\
$^{4}$Netherlands Institute for Radio Astronomy, Oude Hoogeveensedijk 4, 7991 PD Dwingeloo, The Netherlands\\
$^{5}$Astronomical Institute of the Romanian Academy, Cutitul de Argint 5, RO-040557 Bucharest, Romania\\
$^{6}$Research Center for Atomic Physics and Astrophysics, Atomistilor 405, RO-077125 Bucharest, Romania\\
$^{7}$Department of Astronomy, University of Michigan, 500 Church Street, Ann Arbor, MI 48109, USA\\
$^{8}$Kavli Institute for Astrophysics and Space Research, Massachussetts Institute of Technology, Cambridge MA 02139, USA\\ 
$^{9}$INAF-Osservatorio Astronomico di Brera, via E. Bianchi 46, I-23807 Merate (LC), Italy\\ 
$^{10}$NRAO Headquarters, 520 Edgemont Road, Charlottesville, VA 22903, USA\\
$^{11}$Jodrell Bank Centre for Astrophysics, University of Manchester, Oxford Road, Manchester M13 9PL, UK\\
$^{12}$Altran B.V., Hendrik Walaart Sacrestraat 405, 1117 BM Schiphol-Oost, The Netherlands\\
$^{13}$Leiden Observatory, University of Leiden, PO Box 9513, 2300 RA Leiden, The Netherlands\\
$^{14}$Centre for Astrophysics and Supercomputing, Swinburne University of Technology, Hawthorn, Victoria 3122, Australia
}
\begin{document}

\date{Accepted 2010 April 2. Received 2010 March 5; in original form 2009 December 1}

\pagerange{\pageref{firstpage}--\pageref{lastpage}} \pubyear{2010}

\maketitle

\label{firstpage}

\begin{abstract}
In studies of accreting black holes in binary systems, empirical relations have been proposed to quantify the coupling between accretion processes and ejection
mechanisms. These processes are probed respectively by means of X-ray and radio/optical-infrared observations. The relations predict, given certain accretion conditions,
the expected energy output in the form of a jet. We investigated this coupling by studying the black hole candidate Swift J1753.5-0127, via multiwavelength 
coordinated observations over a period of $\sim 4$ years. We present the results of our campaign showing that, all along the outburst, the source features a jet that
is fainter than expected from the empirical correlation between the radio and the X-ray luminosities in hard spectral state. Because the jet is so weak in this
system the near-infrared emission is, unusually for this state and luminosity, dominated by thermal emission from the accretion disc. We briefly discuss the importance
and the implications of a precise determination of both the slope and the normalisation of the correlations, listing some possible parameters that broadband jet models
should take into account to explain the population of sources characterized by a dim jet. We also investigate whether our data can give any hint about the nature of the
compact object in the system, since its mass has not been dynamically measured.
\end{abstract}


\section{Introduction}  \label{par:intro}
Transient black hole candidates (BHCs) are low-mass X-ray binaries that usually show relatively short (weeks to months)
outbursts, separated by long periods of quiescence (see Remillard \& McClintock 2006 for a review).
When an outburst starts, BHCs go through a loop in an X-ray hardness-intensity diagram (HID),
in which in many cases they draw a q-shape pattern.
Four spectral states (Homan et al. 2001, Homan \& Belloni 2005, Belloni 2009; see McClintock \& Remillard 2006 for an alternative definition of states)
can be identified in the X-ray HID: the low/hard state (LHS), the high/soft state (HSS) and two intermediate states (called hard and soft intermediate state).
The LHS of BHCs is characterized by a hard power-law X-ray spectrum (with photon index $\Gamma \sim 1.4-2.1$;
e.g. Remillard \& McClintock 2006) which is usually interpreted as the result of comptonization of seed photons by hot
electrons (the so called ``corona'', e.g. Esin et al. 1997). Observations suggest that the accretion disc
is cold and truncated at large radii (e.g. McClintock et al. 2001, Tomsick et al. 2009, Done \& Diaz Trigo 2009; but see also Miller et al. 2006b and Reis et al. 2010)
at low X-ray luminosities ($L_{X} \lesssim 1 \% L_{Edd}$ where $L_{Edd}$ is the Eddington luminosity; Cabanac et al. 2009). The X-ray energy 
spectrum of BHCs in the HSS is dominated by a thermal emission below $\sim 5$ keV, likely produced by an optically
thick/geometrically thin accretion disc extending to/close to the innermost stable circular orbit (ISCO). The energy spectrum in the HSS also
features a steeper power-law tail than in the LHS.
The intermediate states are characterized by spectral properties in between the LHS and the HSS. In timing studies,
the LHS and the intermediate states show strong quasi periodic oscillations (QPOs) and noise components while the HSS is
characterized by weak/absent variability (see van der Klis 2006 and Belloni 2009 for reviews).
\begin{figure*}
\begin{tabular}{c}
\resizebox{15cm}{!}{\includegraphics{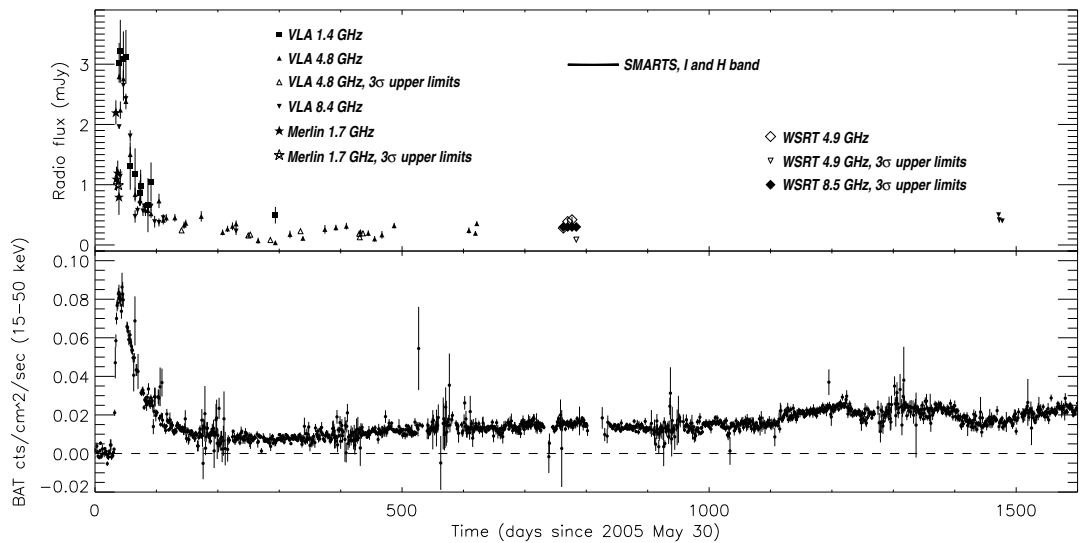}}
\end{tabular}
\caption{Lower panel: Swift/BAT light curve of Swift J1753.5-0127 for the period 2005 May 30 - 2009 October 16. Bin size is 1 day. The dashed line
marks the zero of the {\it y} axis. Upper panel: radio light curve of Swift J1753.5-0127. We plotted all our data points (VLA, MERLIN and WSRT).
Bin size is a whole observation. The horizontal thick line marks the period in which we performed SMARTS OIR observations (H and I band). A complete log of all the radio
and OIR observations is reported in the appendix (\S \ref{app_tables}).}
\label{fig:BAT_radio_SMARTS}
\end{figure*}

A characterization of the four states needs to take into account the behaviour at longer wavelengths (e.g. Fender, Belloni \& Gallo 2004; Fender, Homan \& Belloni
2009).
In the LHS a compact, steady jet is on and substantially contributes to the total energy output of the system (see Fender
2006 for a review). The key signature of such a jet is a flat/slightly inverted spectrum with spectral index $\alpha \gtrsim 0$ ($S_{\nu} \propto \nu^{\alpha}$, where
$S_{\nu}$ is the radio flux density at a certain frequency $\nu$) observed in the
radio band (e.g. Hjellming \& Wade 1971) and extending from radio to higher (sometimes to optical/infrared, OIR) frequencies (Markoff et al. 2003). Following a classical
argument presented in Blandford \& K\"onigl (1979), the jet spectrum can be attributed to a superposition of self-absorbed synchrotron spectra from segments of a
collimated jet. In the last decade jet-dominated models have been developed (Falcke \& Biermann 1995, Markoff, Falcke \& Fender 2001, Giannios 2005, Maitra et al. 2009,
Pe'er \& Casella 2009; but see also Zdziarski et al. 2003, Heinz 2004, Maccarone 2005) and have been used to successfully fit broadband spectra of BHCs in the LHS and in
quiescence (e.g. Gallo et al. 2007; Migliari et al. 2007). In this framework the comptonizing medium (the corona) responsible for the hard X-ray power-law tail
could actually be the base of the jet (Markoff, Nowak \& Wilms 2005; but see also Malzac \& Belmont 2009). There is evidence that the compact jet quenches when the BHC switches to the soft states (Tananbaum et al. 1972; Fender et al. 1999). Intermittent radio emission (characterized by an optically thin spectrum with spectral
index $\alpha < 0$) is occasionally detected in the HSS (Fender et al. 2009) and it could be attributed to the interaction between ejecta detached from the jet base and the interstellar medium. However, the possibility that it originates in a weaker core jet itself can not be completely ruled out without high angular resolution VLBI
(Very Long Baseline Interferometry) radio observations.

Not all the BHCs undergo outbursts drawing a q-shape path in the HID: GRS 1915+105 for example spends all its time in the intermediate states at
very high X-ray luminosity (see Fender \& Belloni 2004 for a review), while there is a number of sources that stay for the whole outburst in the LHS
(Brocksopp, Bandyopadhyay \& Fender 2004) or (possibly) in the LHS and in the hard-intermediate state (SAX J1711.6-3808, Wijnands \& Miller 2002), without transiting to the soft states. Some of these systems underwent both ``normal'' outburst and
LHS-only outbursts (e.g. XTE J1550-564, Homan et al. 2001, Belloni et al. 2002). H~1743-322 is the only known BHC that exhibited both ``normal'' outbursts (see e.g. Jonker et al. 2010) and one outburst in the LHS and in the hard-intermediate state only (Capitanio et al. 2009). The possibility of explaining both type of outbursts
(with/without a transition to the soft states) is a challenge for theoretical models: both the disc-instability model (DIM, King \& Ritter 1998,
Dubus et al. 2001, Lasota 2001) and the ``diffusive'' model (Wood et al. 2001) expect the outburst to take place in the disc, but contribution
from a corona and/or a jet needs to be added to explain the LHS emission and possibly the quiescent one, if the jet indeed plays a substantial
role (e.g. Gallo et al. 2006).

To understand the mechanisms governing BHC outbursts it is important to quantify the contribution of the different processes (accretion/ejection) to
the total energy output of the system. Hannikainen et al. (1998) and Corbel et al. (2003) found an empirical correlation between the radio flux density and the
X-ray flux in the LHS of the BHC GX 339-4.
Gallo, Fender \& Pooley (2003) enlarged the sample including other sources in the LHS, covering more than three orders of magnitude in X-ray luminosity and proposed
that the correlation could be universal in both slope and normalisation. Since the radio emission very likely originates in the compact jet and the X-ray emission
from the accretion processes, assuming that the two are causally related, then the correlation constitutes a physical relation between the inflow and the outflow.
The contribution of the compact jet to the total energy output of BHCs in the LHS sometimes extends to OIR frequencies (e.g. Markoff et al. 2003; in the optical
band the emission is dominated by the companion star and the heated/irradiated accretion disc rather than the jet).
Russell et al. (2006) tested whether a correlation between the X-ray flux and the OIR flux also holds, in the LHS and in quiescence. They found that a global
correlation exists, extending over $\sim$9 orders of magnitude in X-ray luminosity (2-10 keV).\\ 
Gallo et al. (2006) included in the sample data from the BHC A0620-00 in quiescence. In this way they extended the radio/X-ray correlation from
typical LHS levels down to very low X-ray luminosities ($L_X \sim 10^{-8.5} L_{Edd}$ for a distance to the source of 1.2 kpc), showing that a jet can still be
detected at very low accretion rates (with a radio flux at $\mu$Jy level at 8.5 GHz), suggesting that the physics of the inflow/outflow coupling in the LHS probably
still holds in quiescence.
The BHC V404 Cyg also produces a jet when in its quiescent state (Gallo et al. 2003; Miller-Jones et al. 2008), although its quiescent X-ray luminosity is rather
high ($L_{X} \sim 5.3 \times 10^{32}$ erg/s, Corbel, K\"{o}rding \& Kaaret 2008; we considered a distance to the source
of $2.39\pm0.14$ kpc, Miller-Jones et al. 2009), as expected (Lasota 2008) for a long orbital period BHC ($\sim 155$ hours).

The same X-ray/radio scaling found for BHCs may hold for supermassive black holes in active galactic nuclei (AGN), if the mass of the black hole is
taken into account. Merloni, Heinz \& di Matteo (2003) and Falcke, K\"{o}rding \& Markoff (2004) have independently shown that BHCs and AGN populate
a ``Fundamental Plane'' (FP) in the log($L_R, L_X, M$) domain. This suggests that the same mechanisms govern accretion and ejection processes from black holes
hold over $\sim$ 9 orders of magnitude in mass.\\
K\"{o}rding et al. (2006b) noted that, since the accretion states of BHCs are defined according to the source position on an X-ray HID, a generalization
of HID could be constructed also for AGN. In this way they showed that radio loud BHCs and AGN populate the same region of a HID, suggesting that
despite the different masses involved, systems that contain a black hole display similar accretion states and jet properties.\\
McHardy et al. (2006) showed that BHCs in the HSS and AGN also populate a plane in the space defined by the black hole mass $M$, the accretion rate $\dot{M}$ and
a characteristic frequency $\nu$ in the X-ray power density spectra. The plane has been extended by K\"{o}rding et al. (2007) by also including BHCs in the
LHS (considering a constant offset for the frequencies in the two states). This also suggests that some fundamental properties of BHCs and AGN, like
characteristic timescales and mass accretion rate, are related in the same way in the two classes of objects, once the mass has been taken into account.  

The existence of the radio/X-ray correlation and the FP have broad implications. For example, the small scatter around them has been used as
an argument by Heinz \& Merloni (2004) to infer that jets from BHCs and AGN (once a mass-correction factor is introduced) are characterized by
similar bulk velocities. However, in the last years, the supposed universality of the radio/X-ray correlation has been doubted (Xue \& Cui 2007) and several outliers have
been found (Gallo 2007). These sources seem to follow ``normal'' outbursts in the X-rays (their X-ray luminosities are similar to the other BHCs) but they are fainter
in radio (at the same X-ray luminosity) than other sources. Corbel et al. (2004) and Gallo (2007) proposed that a correlation with the same slope but a lower normalisation
(a factor $\sim 20$) could describe this discrepancy, at least in a few sources (e.g. in the BHC XTE J1650-500, Corbel et al. 2004). How BHCs accreting at similar
rates (and so displaying similar X-ray luminosities) can produce different ejecta has not been clarified yet. Considering the similarities in the inflow/outflow coupling
between BHCs and AGN, any explanation should also be relevant for supermassive black holes (possibly as far as helping to explain the apparent radio loud:radio quiet
dichotomy, e.g. Sikora, Stawarz \& Lasota 2007, Tchekhovskoy et al. 2010).

\subsection{Swift J1753.5-0127} \label{par_1753}
Swift J1753.5-0127 was discovered in the hard X-ray band with the Burst Alert Telescope (BAT) on board Swift on 2005 May 30 (Palmer
et al. 2005). The source was also detected in the soft X-rays with the Swift/X-ray Telescope (Swift/XRT, Burrows et al. 2005)
and with the Proportional Counter Array (PCA) aboard the Rossi X-ray Timing Explorer (RXTE, Morgan et al. 2005).
Figure \ref{fig:BAT_radio_SMARTS} (bottom panel) shows the Swift/BAT light curve: after the discovery, the source flux
reached a peak of 200 mCrab on 2005 July 9 in the All Sky Monitor (ASM, 1.2-12 keV) on board RXTE (Cadolle Bel et al. 2007). Subsequently the
flux started decreasing and then stalled at a level of $\sim$20 mCrab (2-20 keV) for more than $\sim$ 6 months. This is an
unusual behaviour for a transient, but even more unusual is the subsequent flux rise which has been ongoing since roughly June 2006, with a steepening
in June 2008 (Krimm et al. 2008), followed by a decreasing/variable trend. No transition to the soft states has
been reported: the source has always been in the LHS during the whole outburst (Zhang et al. 2007; Cadolle Bel et al. 2007), although
its X-ray spectral hardness has not remained constant (Zhang et al. 2007, Ramadevi \& Seetha 2007, Negoro et al. 2009).
The source was also detected in ultraviolet with the Ultraviolet/Optical telescope UVOT aboard Swift (Still et al. 2005) and in optical with the Michigan-Dartmouth-MIT
2.4 m telescope (Halpern 2005). In July 2005, near the peak of its outburst, Fender, Garrington \& Muxlow (2005) observed Swift J1753.5-0127 at radio frequencies with the
Multi-Element Radio-Linked Interferometer Network (MERLIN), tentatively detecting a point-like radio counterpart, consistent with the presence of a compact jet.\\
The strongest hint that the system harbours a black hole comes from the hard power-law tail in the X/$\gamma$-ray energy spectrum up to $\sim$600 keV, detected with INTEGRAL (Cadolle Bel et al. 2007), as no low-mass neutron star X-ray binary (NSXB) has ever been detected above $\sim 200$ keV (di Salvo et al. 2006; Falanga et al. 2007). To date four
(or possibly five) high-mass X-ray binaries have been detected at TeV energies (see e.g. Rea et al. 2010 and references therein), although the nature of the accretor is clear only in two of them (Cyg X-1 contains a black hole and PSR B1259-63 harbours a neutron star; see Paredes \& Zabalza 2010). In all these sources, the spectrum at TeV energies is not compatible with the extrapolation of the hard X-ray spectrum (see e.g. Sidoli et al. 2006 for LS I +61 303).\\ 
The mass of the compact object in Swift J1753.5-0127 has not been dynamically measured. Bearing all these considerations in mind, we will treat Swift J1753.5-0127 as a black hole candidate.\\ 
Cadolle Bel et al. (2007) considered both the high Galactic latitude of Swift J1753.5-0127 and the low Galactic column density in its direction and
concluded that its distance should not be larger than 10 kpc, most likely in the $4-8$ kpc
range. Using an empirical relation that predicts the absolute magnitude of the accretion disc of a BHC in outburst (from Shahbaz \& Kuulkers 1998),
Zurita et al. (2008) derived a distance to the source $D > 7.2 kpc$. Throughout this paper we will consider a distance $D = 8$ kpc to Swift J1753.5-0127, unless it
will be differently specified.    

Swift J1753.5-0127 is peculiar for a number of reasons. First of all it is an outlier to the radio/X-ray luminosity correlation of Gallo et al. (2003). According
to Cadolle Bel et al. (2007), Swift J1753.5-0127 should be 8-60 times more luminous in radio (depending on the distance,
assuming that it must be inside our Galaxy), to fit the correlation. This BHC never left the LHS, so we expect its jet to
substantially contribute to the total energy budget of the system, as observed in most of the BHCs in the LHS.\\
Secondly, it would be the BHC with the shortest orbital period, as claimed by Zurita et al. (2008) and Durant et al. (2009), who reported
a $\sim$3.2 hr modulation in the optical lightcurves. It is worth to note (Zurita et al. 2008) that such period, if confirmed, would be close
to the orbital period of two other BHCs observed only in the LHS, XTE J1118+480 and GRO J0422+32: they respectively feature orbital periods of 4.1 hr
(McClintock et al. 2001; Wagner et al. 2001) and 5.1 hr (Filippenko et al. 1995). Interestingly, all these sources are likely located in the
Galactic halo (Cadolle Bel et al 2007, Zurita et al. 2008, Durant et al. 2009) and they could constitute a population of high Galactic latitude X-ray binaries,
putting constraints on their formation and evolution. Hynes et al. (2009) recently performed coordinated optical and X-ray observations of Swift
J1753.5-0127 at high time resolution, reporting a short delay between the optical and the X-ray emission, consistent with an orbital period as short as
3.2 hours.\\
From binary evolution calculations nothing prevents BHCs to have in principle orbital periods of $\sim$2 hours and evolutionary
models actually predict that short-period systems might form the majority of them (Yungelson et al. 2006). Nevertheless, such a short period would
definitely be peculiar for a BHC (see e.g. Charles \& Coe 2006). NSXBs usually feature a fainter
jet than BHCs (about a factor $\sim$30 in radio flux; Fender \& Hendry 2000; Migliari \& Fender 2006) at the same X-ray luminosity. This could suggest
that the compact object in the system is a neutron star and not a black hole, although other evidence points towards a black-hole nature, as mentioned above.

Another interesting aspect concerning Swift J1753.5-0127 is related to the possible presence of an accretion disc extending to the
ISCO even if in the LHS (Miller et al. 2006a), at variance with the ``standard'' picture in which the disc is truncated
at larger radii (e.g. Esin et al. 1997) in the LHS at low luminosities (Cabanac et al. 2009; Tomsick et al. 2009). The need for a thermal component extending to
the ISCO to fit the source spectra has been recently weakened by Hiemstra et al. (2009): they have shown that several spectral models,
without necessarily including disc components extending to the ISCO, can fit the data equally well (but see Reis et al. 2009 and Reynolds et al. 2010).

In this paper we present the results of radio, OIR and X-ray observations of Swift J1753.5-0127 performed from the beginning of the outburst
(2005 July) until 2009 June (\S\ref{par:data}). During this interval, we collected a large sample of (quasi) simultaneous radio/X-ray and
OIR/X-ray observations to test whether the jet is systematically less luminous than predicted by the empirical correlations introduced
above (\S \ref{par:gallo_rel} and \S \ref{par:russell_rel}). In 2007 July we also obtained (quasi) simultaneous multiwavelength observations from radio up to hard X-ray
frequencies, that we used to produce spectral energy distributions (SEDs, \S \ref{par:SED}). Our results will be presented in \S \ref{par:results} and discussed in
\S \ref{par:discussion}. In the last section we will summarize our conclusions (\S \ref{par:conclusions}).

\section{Observations and data analysis} \label{par:data}
We observed Swift J1753.5-0127 from the peak of its outburst (2005 July) until 2009 June. In this
section we will describe the data reduction performed, for all the instruments used.
\begin{table*}
\centering
\caption{Log of the RXTE and Swift observations analysed in this paper. We report the spectral model used to get the X-ray flux to test the
radio/X-ray (2-11 keV) and the OIR/X-ray correlation (2-10 keV). A multiplicative constant has always been added to the model in order
to allow normalisation between the different instruments. We also show the values of the best-fit $\chi^2$. We report a key to the models: phabs stands for
photolectric absorption; bknpower for broken power law, diskbb for disc blackbody and pow for power law.}
\label{tab:log_obs_X}
\begin{tabular}{c c c c c c}
\hline
\hline
\multicolumn{6}{c}{{\bf Radio/X-ray correlation}}\\
Date        & RXTE Obs.ID    & Swift Obs.ID & Model                           & $\chi^{2}/dof^{\mathrm{e}}$ & 2-11 keV flux$^{\mathrm{a}}$   \\ 
\hline
2005-07-04  & 91094-01-01-01 &       -      & phabs bknpower diskbb           & 62.61/59       & $4.63_{-0.05}^{+0.01}$         \\
2005-07-06  & 91423-01-01-04 &  00030090003 & phabs pow diskbb$^{\mathrm{b}}$ & 1007/656       & $3.70_{-0.50}^{+0.84}$           \\ 
2005-07-07  & 91423-01-01-00 &       -      & phabs bknpower diskbb           & 69.34/59       & $4.861_{-0.058}^{+0.006}$        \\
2005-07-08  & 91094-01-02-01 &  00030090007 & phabs pow$^{\mathrm{b,c}}$      & 796/693        & $3.79_{-0.63}^{+0.91}$           \\
2005-07-10  & 91094-01-02-02 &  00030090011 & phabs bknpower diskbb$^{\mathrm{b}}$ & 495/368   & $3.61_{-0.03}^{+0.07}$         \\
2005-07-19  & 91423-01-03-04 &       -      & phabs bknpower diskbb           & 57.78/59       & $3.588_{-0.030}^{+0.005}$        \\
2005-07-26  & 91423-01-04-04 &       -      & phabs bknpower diskbb           & 45.45/59       & $2.850_{-0.023}^{+0.007}$        \\
2005-08-03  & 91423-01-05-02 &       -      & phabs bknpower diskbb           & 58.07/59       & $2.119_{-0.024}^{+0.001}$        \\
2005-08-07  & 91423-01-06-01 &       -      & phabs bknpower diskbb           & 45.79/59       & $1.802_{-0.018}^{+0.005}$        \\
2005-08-11  & 91423-01-06-03 &       -      & phabs pow diskbb                & 87.42/61       & $1.670_{-0.026}^{+0.005}$        \\ 
2005-09-11  &       -        &  00030090024 & phabs pow                       & 1297/1242      & $0.813_{-0.017}^{+0.004}$        \\
2005-10-22  & 91423-01-17-00 &  00030090031 & phabs pow                       & 677/528        & $0.559_{-0.002}^{+0.003}$       \\
2005-11-19  & 91423-01-21-00 &       -      & phabs pow                       & 57.4/63        & $0.464_{-0.003}^{+0.002}$       \\ 
2006-03-11  & 92404-01-02-00 &       -      & phabs pow                       & 55.1/63        & $0.383_{-0.002}^{+0.003}$       \\
2006-08-03  &       -        &  00030090032 & phabs pow                       & 331/365        & $0.434_{-0.014}^{+0.007}$        \\
2007-07-01  & 93105-01-08-00 &  00030090042 & phabs pow                       & 765/732        & $0.742_{-0.001}^{+0.012}$        \\
2007-07-08  & 93105-01-09-00 &  00030090045 & phabs pow                       & 1055/1001      & $0.72_{-0.12}^{+0.11}$           \\
2007-07-15  & 93105-01-10-00 &  00030090050 & phabs pow                       & 830/747        & $0.766\pm0.009$                \\
2007-07-22  & 93105-01-11-00 &       -      & phabs pow                       & 51.73/60       & $0.676_{-0.005}^{+0.002}$       \\
2009-06-09  & 93105-02-33-00 &       -      & phabs pow                       & 45.48/50       & $0.731_{-0.003}^{+0.002}$       \\
\hline 
\multicolumn{6}{c}{{\bf OIR/X-ray correlation}}\\
Date        & RXTE Obs.ID    & Swift Obs.ID & Model                           & $\chi^{2}/dof$ & 2-10 keV flux$^{\mathrm{d}}$   \\ 
\hline 
2007-07-08  & 93105-01-09-00 &  00030090045 & phabs pow 		      & 1055/1001      & $6.71_{-0.58}^{+1.17}$		\\
2007-07-15  & 93105-01-10-00 &  00030090050 & phabs pow 		      & 830/747        & $7.08_{-0.06}^{+0.11}$  	\\
2007-07-22  & 93105-01-11-00 &       -      & phabs pow 		      & 51.73/60       & $6.25 \pm 0.03$		\\
2007-07-29  & 93105-01-12-00 &       -      & phabs pow diskbb                & 45.22/58       & $6.77_{-0.14}^{+0.02}$          \\
2007-08-05  & 93105-01-13-00 &       -      & phabs pow diskbb                & 56.97/58       & $6.58_{-0.08}^{+0.01}$         \\
2007-08-12  & 93105-01-14-00 &       -      & phabs pow diskbb                & 56.81/58       & $5.82_{-0.09}^{+0.03}$         \\
2007-08-20  & 93105-01-15-00 &       -      & phabs pow                       & 58.58/60       & $5.8856_{-0.0420}^{+0.0007}$   \\
2007-09-09  & 93105-01-18-00 &       -      & phabs pow                       & 69.44/60       & $5.56_{-0.05}^{+0.01}$         \\
2007-09-23  & 93105-01-20-00 &       -      & phabs pow diskbb                & 57.40/58       & $5.67_{-0.22}^{+0.01}$         \\
\hline 
\hline
\end{tabular}
\begin{list}{}{}
\item[$^{\mathrm{a}}$] In $10^{-9}$ erg/cm$^2$/s, un-absorbed flux
\item[$^{\mathrm{b}}$] Swift/XRT spectra fitted only in the 0.5-10 keV band
\item[$^{\mathrm{c}}$] RXTE/HEXTE data not available for this observation
\item[$^{\mathrm{d}}$] In $10^{-10}$ erg/cm$^2$/s, un-absorbed flux
\item[$^{\mathrm{e}}$] dof: degrees of freedom
\end{list}
\end{table*}

\subsection{X-ray data} \label{par:X-ray_data}
\subsubsection{RXTE}
\begin{figure*}
\begin{tabular}{c}
\resizebox{16.8cm}{!}{\includegraphics{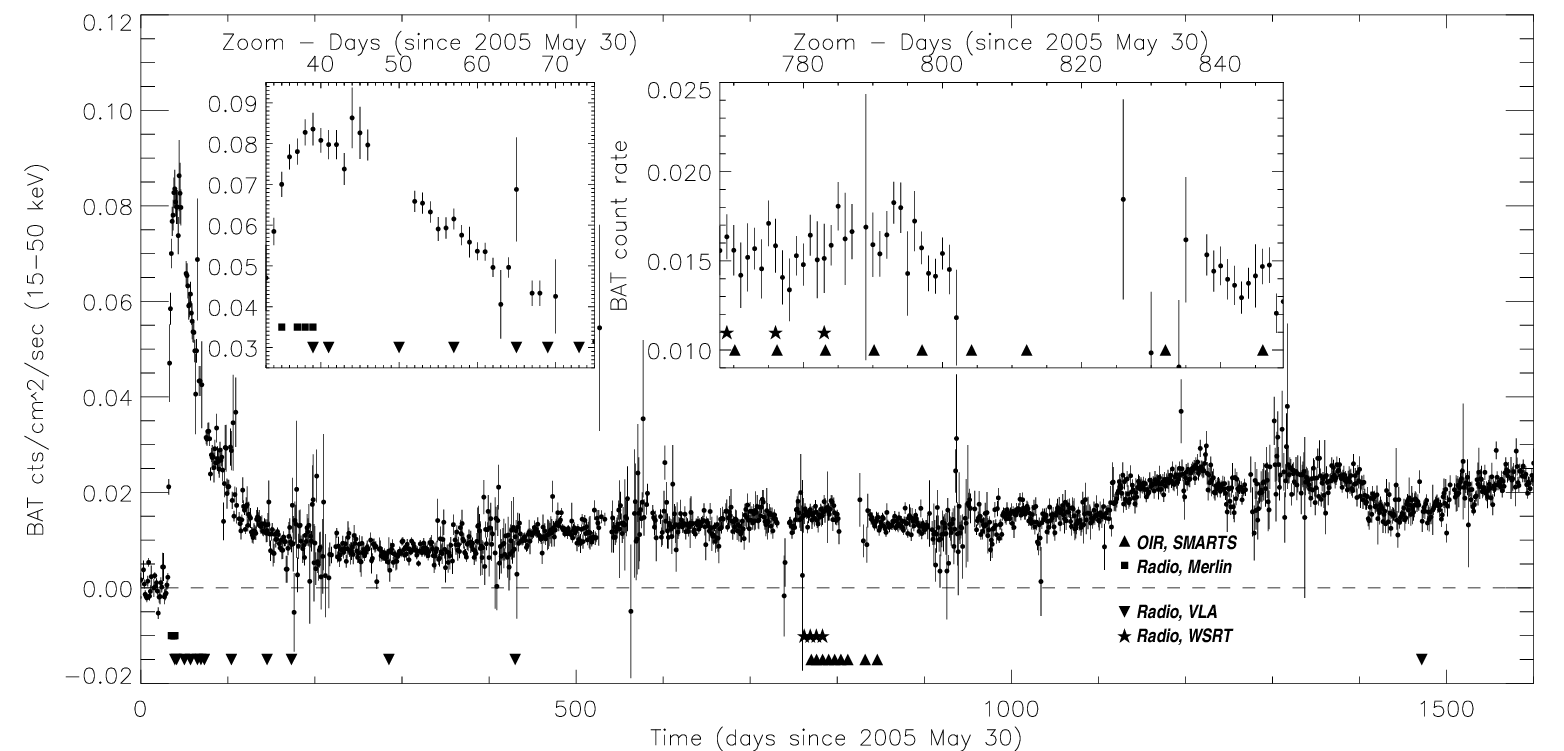}}
\end{tabular}
\caption{Swift/BAT light curve of Swift J1753.5-0127, as in Figure \ref{fig:BAT_radio_SMARTS}. Different symbols at the lower edge of the figure represent the
radio (VLA, MERLIN and WSRT) and SMARTS observations listed in Tables \ref{tab:log_obs_radio} and \ref{tab:log_obs_smarts}. For clarity, the insets show a zoom on
two intervals (2005 July 02 - 2005 August 13 and 2007 July 07 - 2007 August 26).}
\label{fig:BAT_licu}
\end{figure*}
We analysed 24 RXTE observations of Swift J1753.5-0127 (Table \ref{tab:log_obs_X}). We applied standard filters and
extracted PCA spectra from Standard 2 data (Jahoda et al. 1996), using only data from PCU2 (Proportional Counter Unit 2), the most reliable of the
five PCUs, always on during all our observations. The data were background corrected using the latest PCA background model. We generated response
matrices and fitted the spectra in the energy range 3.0-25.0 keV, after applying a 1\% systematic error. We also extracted background and dead-time
corrected HEXTE (High Energy Timing Experiment on board RXTE, cluster B only) energy spectra that we fitted in the energy range 20-200 keV, after
generating response matrices. HEXTE spectra in the energy channels 1-26 were grouped by a factor 2, while in the channels 27-62 they were grouped
by a factor 3. No systematic error was applied. To produce the SEDs discussed in \S \ref{par:SED} we used the HEXTE spectra in the energy range 20-150 keV.

\subsubsection{Swift}
We ran the XRT pipeline (v.0.12.0) on all the 9 Swift observations (see Table \ref{tab:log_obs_X}), using standard quality cuts and event grades
0-2. Our data were all collected in windowed timing (WT) mode. Since within each observation there are time gaps in the data, we obtained an event
file for every observing interval. We extracted source and background spectra (with Xselect v.2.3) for every event file using circular regions
with radii of $\sim$30 and $\sim$15 pixel, respectively. We generated exposure maps and we used them as an input to create ancillary response files.
We made use of the latest response matrices (v011) provided by the Swift/XRT team. The spectra have been grouped to a minimum of 20 counts per bin.
A systematic error of 2.5\% was also applied (Campana et al. 2008). We fitted the spectra in the 0.3-10 keV energy range. For three
observation we could get a statistically acceptable fit only by restricting the energy range to 0.5-10 keV (see Table \ref{tab:log_obs_X}).
To produce the SEDs discussed in \S \ref{par:SED} we used the XRT spectra in the energy range 0.6-10 keV.

\subsubsection{Fitting procedures}
We fitted the X-ray spectra using the standard XSPEC v11.3 fitting package (Arnaud 1996). For simultaneous RXTE and Swift observations
(performed within the same day) we combined the XRT+PCA+HEXTE spectra to fit them together. For a normalisation between the different
instruments, we multiplied the continuum model by a constant component. All model parameters, except these multiplicative constants, were linked
between the different instruments. When fitting RXTE spectra alone, we fixed the interstellar absorption to the average column density value
obtained by Hiemstra et al. 2009 ($N_H = 1.7 \times 10^{21}$cm$^{-2}$) while we let it free to vary when including also XRT data.
We fitted Swift J1753.5-0127 energy spectra in order to get a reliable estimate of the un-absorbed X-ray flux in two energy bands (2-10 keV and 2-11 keV, to make the
comparison easier with the results in Russell et al. 2006 and Gallo et al. 2006, respectively, see \S \ref{par:results}), so an accurate discussion on the best
model to be used is beyond the scope of this paper. For that we address the reader to Miller et al. (2006a), Reis et al. (2009), Hiemstra et al. (2009), Cabanac
et al. (2009) and Reynolds et al. (2010). The models used for the fitting are reported in Table \ref{tab:log_obs_X}.
We obtained reduced $\chi^{2}$ between 0.86 and 1.53. 

\subsection{Radio data} \label{par:radio_data}
\begin{table*}
\centering
\caption{Log of the MERLIN, VLA (proposal codes AR570, AR572, AR603 and AM986) and WSRT observations simultaneous (performed within 24 hours) to the RXTE
and/or Swift observations listed in Table \ref{tab:log_obs_X}. We used these observations to test the radio/X-ray correlation in
Figure \ref{fig:Gallo_relation}. The data points plotted in Figure \ref{fig:BAT_radio_SMARTS} (upper panel) that are not simultaneous to any pointed
RXTE and/or Swift observation are not listed here. Figure \ref{fig:BAT_licu} shows the position of these radio observation on the Swift/BAT light curve
of the outburst of Swift J1753.5-0127.}
\label{tab:log_obs_radio}
\begin{tabular}{c c c c c c c c}
\hline
\hline
\multicolumn{8}{c}{{\bf Radio observations}}\\
           & \multicolumn{6}{c}{Flux densities (mJy)}                                                                             & Spectral index $\alpha$ \\
           &     MERLIN          &                    \multicolumn{3}{c}{VLA}                   &       \multicolumn{2}{c}{WSRT}  &                         \\
Date       &    1.7 GHz          &     1.4 GHz   &     4.8 GHz              &      8.4 GHz      &   4.9 GHz           &   8.5 GHz &                         \\ 
\hline
2005-07-04 & $1.1\pm0.2$	 &	-	 &	-		    &	    -	        &      - 	      &      -	  &           -             \\ 
2005-07-06 & $1.2\pm0.2$	 &	-	 &	-		    &	    -	        &      - 	      &      -	  &	      -             \\
2005-07-07 & $1.0^{\mathrm{a}}$  &	-	 &	-		    &	    -	        &      - 	      &      -	  & 	      - 	    \\
2005-07-08 & $0.8\pm0.3$	 & $3.02\pm0.33$ & $2.79\pm0.05$            & $1.96\pm0.04$     &      - 	      &      -	  &	$-0.30\pm0.27$      \\
2005-07-10 &	-		 & $3.21\pm0.52$ & $2.24\pm0.14$	    & $1.14\pm0.12$     &      - 	      &      -	  &	$-0.59\pm0.30$      \\
2005-07-19 &	-		 & $3.12\pm0.45$ & $2.38\pm0.13$	    & $2.42\pm0.05$     &      - 	      &      -	  &	$-0.12\pm0.37$      \\
2005-07-26 &	-		 & $1.31\pm0.39$ & $1.50\pm0.13$	    & $1.81\pm0.09$     &      - 	      &      -	  &	$0.26\pm0.09$       \\
2005-08-03 &	-		 & $1.18\pm0.43$ & $0.84\pm0.12$	    & $0.47\pm0.10$	&      - 	      &      -	  &     $-0.42\pm0.31$      \\
2005-08-07 &	-		 &	 -	 &	-		    & $0.57\pm0.07$     &      - 	      &      -	  &	      -             \\
2005-08-11$^{\mathrm{b,c}}$  &  -  & $0.86\pm0.21$ & $0.74\pm0.05$	    & $0.68\pm0.01$	&      -	      &      -	  &     $-0.14\pm0.01$      \\
2005-09-11 &	 -		 &	 -	 & $0.73\pm0.12$	    & $0.37\pm0.07$     &      -	      &      -    &     $-1.20^{\mathrm{d}}$          \\ 
2005-10-22 &	 -		 &	 -	 & $0.33\pm0.06$	    &       -  		&      -	      &      -    &	      -	            \\
2005-11-19 &	 -		 &	 -	 & $0.48\pm0.09$	    &       -  		&      -	      &      -    &	      -  	    \\ 
2006-03-11 &	 -		 &	 -	 & $0.08^{\mathrm{a}}$      &       -  		&      -	      &      -    &	      - 	    \\
2006-08-03 &	 -		 &	 -	 & $0.19^{\mathrm{a}}$      &       -  		&      -	      &      -    &	      - 	    \\
2007-07-01 &	 -		 &	 -	 &	 -		    &       -  		& $0.28\pm0.05$       & $0.3^{\mathrm{a}}$ & $\lesssim 0.13$ \\
2007-07-08 &	 -		 &	 -       &	 -		    &       -  		& $0.39\pm0.05$       & $0.3^{\mathrm{a}}$ & $\lesssim -0.48$ \\
2007-07-15 &	 -		 &	 -	 &	 -		    &       -  		& $0.42\pm0.05$       & $0.3^{\mathrm{a}}$ & $\lesssim -0.61$ \\
2007-07-22 &	 -	         &	 -       &	 -		    &       -  		& $0.09^{\mathrm{a}}$ & $0.3^{\mathrm{a}}$ &   -            \\
2009-06-09 &	 -		 &	 -       &	 -		    & $0.50\pm0.02$	&      -              &      -    &	       -            \\ 
\hline 
\hline
\end{tabular}
\begin{list}{}{}
\item[$^{\mathrm{a}}$] $3\sigma$ upper limit
\item[$^{\mathrm{b}}$] 15 GHz flux (VLA): $0.69\pm0.17$ mJy
\item[$^{\mathrm{c}}$] the flux densities that we obtained for this observation are different from the ones quoted in Cadolle Bel et al. (2007), although consistent within
the errors. Probably Cadolle Bel et al. (2007) used slightly different settings for the data reduction and the fitting.
\item[$^{\mathrm{d}}$] the errors on the slope could not be obtained, since we fitted two data point using a model with two free parameters.
\end{list}
\end{table*}
We observed Swift J1753.5-0127 in radio with the Westerbork Synthesis Radio Telescope (WSRT), the Very Large Array (VLA) and MERLIN. A radio light curve is
plotted in Figure \ref{fig:BAT_radio_SMARTS} (upper panel). A log of the observations used to test the X-ray/radio correlation is reported in Table
\ref{tab:log_obs_radio}. A complete log of all the radio observations is reported in table \ref{tab:log_obs_radio_ALL} in the appendix.\\
In the following we will describe the data-reduction steps that we performed for each radio facility.

\subsubsection{WSRT}
We observed Swift J1753.5-0127 with the WSRT on four occasions: 2007 July 01/02, 08/09, 15/16, and 22/23.
The observations were made between $\sim$16--04 UT, in the frequency-switching mode, at the median frequencies of 4.901 and 8.463 GHz, with a total
bandwidth of 160 MHz. The primary calibrators used were 3C 48 for the first epoch and 3C 286 for the others. The calibration and analysis of the data
were done using \textsc{MIRIAD} (Sault et al. 1995).\\
The resulting beam was very elongated. To improve on the quality of the images, we selected only the data taken
between 18--02 UT and restored the radio maps with a circular Gaussian beam with the full width at half maximum of 10 arcsec at 4.9 GHz and 5 arcsec at 8.5
GHz. These sizes of the restoring beams were chosen to be close to the values of the spatial resolution in the East-West direction at the two
frequencies.\\
When detected, at a position compatible with that reported by Fender et al. (2005) and implicitly by Cadolle Bel et al. (2007), the target was
unresolved. The flux densities quoted here were measured in the image plane. We note that on the last epoch, i.e. 2007 July 22/23, at 4.9 GHz we detected
a compact radio source of about 0.25 mJy flux density, 25 arcsec away from the expected position of Swift J1753.5-0127. This is very likely an artifact.
\begin{table}
\centering
\caption{Log of the SMARTS observations used to test the OIR/X-ray correlation. If on one day we had X-ray observations but not a SMARTS pointing, we averaged (when
possible) the I-band and the H-band fluxes obtained on the previous and the following day. Figure \ref{fig:BAT_licu} shows the position of these SMARTS observation on the
Swift/BAT light curve of the outburst of Swift J1753.5-0127. The observed flux densities have been de-reddened assuming $R_{v} = 3.1$ and $A_{v} = 3.1 \times E_{B-V} =
1.05$ ($E_{B-V} \sim 0.34$, Cadolle Bel et al. 2007).}
\label{tab:log_obs_smarts}
\begin{tabular}{l c c}
\hline
\multicolumn{3}{c}{{\bf SMARTS observations}} \\
\multicolumn{1}{c}{Date}  & \multicolumn{2}{c}{Flux densities (mJy)}                     \\
                          &  I band                  & H band                  \\ 
\hline
2007-07-09                & $1.51\pm0.01$            & $1.09\pm0.24$           \\
2007-07-15                & $1.61\pm0.04$            & $1.16\pm0.26$           \\
2007-07-21$^{\mathrm{a}}$ & $1.46\pm0.09$            & $1.10\pm0.26$           \\
2007-07-23$^{\mathrm{a}}$ & $1.69\pm0.11$            & $1.25\pm0.35$           \\
2007-07-22$^{\mathrm{b}}$ & $1.57\pm0.14$            & $1.18\pm0.43$           \\
2007-07-28$^{\mathrm{a}}$ & $1.66\pm0.02$            & $1.15\pm0.25$           \\
2007-07-30$^{\mathrm{a}}$ & $1.71\pm0.09$            & $1.19\pm0.25$           \\
2007-07-29$^{\mathrm{b}}$ & $1.68\pm0.10$            & $1.17\pm0.35$           \\
2007-08-05                & $1.54\pm0.01$            & $1.23\pm0.50$           \\
2007-08-11$^{\mathrm{a}}$ & $1.43\pm0.21$            &      -                  \\
2007-08-13$^{\mathrm{a}}$ & $1.47\pm0.03$            & $1.02\pm0.22$           \\
2007-08-12$^{\mathrm{c}}$ & $1.45\pm0.22$            &      -                  \\
2007-08-19$^{\mathrm{a}}$ & $1.53\pm0.04$            & $1.05\pm0.19$           \\
2007-08-21$^{\mathrm{a}}$ & $1.56\pm0.04$            & $1.29\pm0.25$           \\
2007-08-20$^{\mathrm{b}}$ & $1.54\pm0.06$            & $1.17\pm0.32$           \\
2007-09-09                & $1.51\pm0.03$            & $1.09\pm0.22$           \\
2007-09-22$^{\mathrm{a}}$ & $1.60\pm0.03$	     & $1.06\pm0.20$	       \\
2007-09-24$^{\mathrm{a}}$ & $1.63\pm0.02$	     & $1.17\pm0.23$	       \\
2007-09-23$^{\mathrm{b}}$ & $1.61\pm0.03$            & $1.12\pm0.30$           \\
\hline 
\end{tabular}
\begin{list}{}{}
\item[$^{\mathrm{a}}$] No RXTE/Swift observation available on this day
\item[$^{\mathrm{b}}$] No SMARTS observation available on this day, average of the previous and the following day
\item[$^{\mathrm{c}}$] No SMARTS observation available on this day, average of the previous and the following day (I band only)
\end{list}
\end{table}

\subsubsection{VLA}
All the publicly available VLA data surrounding the outburst at 1.4, 4.8 and 8.4 GHz were retrieved from the VLA archive (proposal codes AR570,
AR572, AR603, AM986, AT320 and S7810). 
We used a ParselTongue\footnote{ParselTongue is a python interface to the Astronomical Imaging Processing System (AIPS, Greisen 2003)} pipeline procedure
(see Kettenis et al. 2006) to calibrate and image the data. Firstly the data were flagged for errors using the AIPS task 'FLAGR'.
As the observations were taken in various VLA configurations at various hour angles, appropriate flux and phase calibrators were used to calibrate
depending on availability. After calibration, the images were made using a natural weighting scheme (for best point source sensitivity) and lightly
cleaned with 50 iterations. Image self calibration was not incorporated as the signal to noise of the target source was insufficient. 

The images were then processed using the prototype LOFAR (LOw Frequency ARray) transient detection algorithms (Swinbank 2007).
The outburst was easily detectable and the algorithms automatically performed Gaussian fits at the region of interest for all detections above 3$\sigma$. 
Below 3$\sigma$ the map {\it rms} was used to produce an upper limit to the flux densities. As the outburst was unresolved on the longest VLA baseline the peak 
flux density from the Gaussian fit was used in Figure \ref{fig:BAT_radio_SMARTS} to plot the evolution of flux with time. 

\subsubsection{MERLIN}
Monitoring observations of Swift J1753.5-0127 were made with the Jodrell Bank MERLIN imaging array in July 2005 at 1658 MHz with a
bandwidth of 15 MHz in both left and right circular polarization (Thomasson 1986, Fender et al. 2005). The observations were made at
the start of the Summer engineering period and contained varying numbers of MERLIN antennas as individual telescopes were withdrawn for
scheduled maintenance and painting. The observations were full tracks running from 18:30 UT to 02:30 UT on each night between July 3rd and
9th. The observations of July 3rd/4th contained the full MERLIN array of 7 telescopes. Observations between July 4th and 6th contained 5
MERLIN antennas, and observations between July 6th and 8th were made with a subset of 3 MERLIN antennas.

The observations were phase referenced to the nearby point source J1743-03 with a target to phase calibrator cycle of 8:2 minutes. The
flux density scale was set by observations of the bright point source OQ208 and the unvarying but resolved calibration source 3C286.
Assuming a total flux density of 13.661 Jy for 3C286, the flux density of OQ208 was derived to be 1.17 Jy at the time of the observations.

The data were re-weighted to account for the varying telescope sensitivities and then all MERLIN baselines vector averaged at the
position of Swift J1753.5-0127 to derive a single flux density measurement for each full track. 
\begin{figure*}
\begin{tabular}{c}
\resizebox{12cm}{!}{\includegraphics{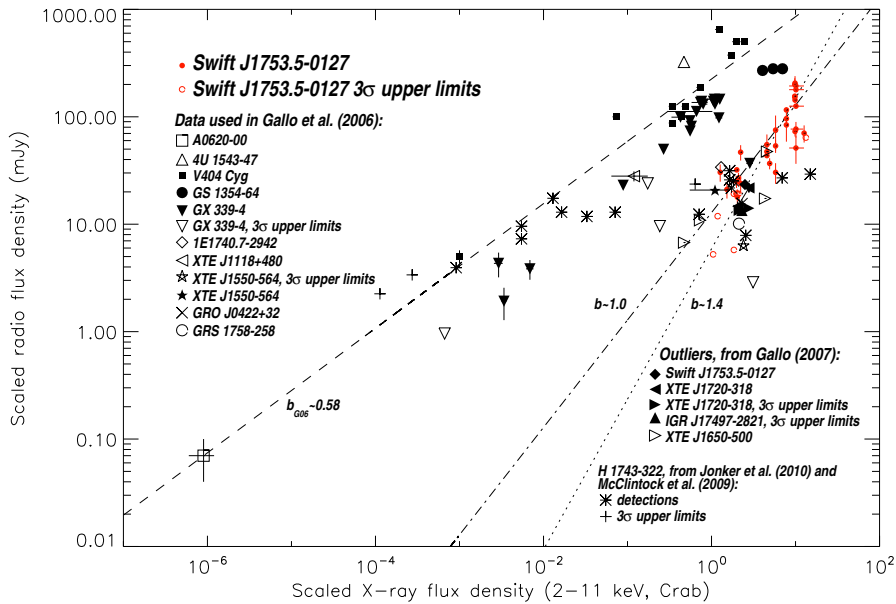}}
\end{tabular}
\caption{The BHCs sample of Gallo et al. (2006) and Gallo (2007), with the addition of the Swift J1753.5-0127 data (red points). We also included data from the
BHC H~1743-322, recently reported  in McClintock et al. (2009) and Jonker et al. (2010). The X-ray scaled fluxes from McClintock et al. (2009) and Jonker et al.
(2010) have been measured in the energy bands 2-20 keV and 0.5-10 keV, respectively. The dashed line is the fit to the Gallo et al. (2006) data, a non-linear
relation of the form $S_{radio} = k (S_{X})^{b}$ ($k_{G06} = 224.72$ and $b_{G06} = 0.58\pm0.16$) where $S_{radio}$ is the scaled radio flux density
and $S_{X}$ is the scaled X-ray flux. The dashed-dotted line and the dotted line represent the best-fit limits to the Swift J1753.5-0127 data, two non linear
relations with $b \sim 1.0$, $k \sim 13$ and $b \sim 1.4$, $k \sim 6$, respectively.
Following Gallo et al. (2003) and Gallo et al. (2006), all the fluxes have been scaled to a distance of 1 kpc.}
\label{fig:Gallo_relation}
\end{figure*}

\subsection{OIR/ultraviolet data}
\subsubsection{SMARTS data} \label{par:smarts_data}
We observed \src\ using the 1.3 m telescope located at Cerro Tololo
Inter-American Observatory (CTIO) with the ANDICAM instrument
(DePoy et al. 2003). ANDICAM
is a dual-channel imager with a dichroic that feeds an optical
CCD and an infrared (IR) imager, which can obtain simultaneous data
in one optical band (B, V, R or I) and one IR band (J, H or K).
Simultaneous I- and H-band images of \src\ were obtained between
2007 July 09 -- 2007 September 23 (this time interval is marked in the upper panel of Figure \ref{fig:BAT_radio_SMARTS}).
Standard flat-fielding and sky subtraction procedures were applied to each
night's data, and the internal dithers in the infrared were combined as described in
Buxton \& Baylin (2004). Differential photometry was carried out each night with a set
of reference stars in the field. Intercomparisons between reference stars of
similar brightness to the source suggest a precision of $<0.02$ mag in I and
$\sim 0.03$ mag in H. Calibrations to the standard optical and IR magnitude
system were carried out using USNO and 2MAHSS stars present in the field of
view of \src.

The observed magnitudes were converted into spectral flux densities using $E(B-V)=0.34\pm0.04$ (Cadolle Bel et al. 2007),
assuming a ratio of total to selective extinction, $R_{v}=3.1$ (Rieke \& Liebofsky 1985), and the extinction law of Cardelli, Clayton \& Mathis (1989).
The uncertainty on the optical extinction dominates over the intrinsic errors (see e.g. Hynes et al. 2002 for a discussion). The de-reddened flux densities obtained
from the SMARTS observations which have been used to test the OIR/X-ray correlation are reported in Table \ref{tab:log_obs_smarts}. A complete log of the SMARTS
observations is reported in the appendix (\S \ref{app_tables}) in Table \ref{tab:OIR_all}.

\subsubsection{Swift/UVOT data} \label{par:uvot_data}
Swift/UVOT observed Swift J1753.5-0127 on 2007 July 08 (Obs.ID 00030090045) with the filters UVW1
(2600\AA), UVM2 (2246\AA) and UVW2 (1928\AA). UVOT observations contain gaps (as the XRT ones), so for each filter we added all the intervals
in order to get one event file per filter. Using the standard HEASOFT tasks we extracted the following fluxes: $(1.02\pm0.03)\times10^{-15}$ 
erg cm$^{-2}$ s$^{-1}$ \AA$^{-1}$ (UVW1), $(7.63\pm0.32)\times10^{-16}$ erg cm$^{-2}$ s$^{-1}$ \AA$^{-1}$ (UVM2) and $(1.20\pm0.03)\times10^{-15}$
erg cm$^{-2}$ s$^{-1}$ \AA$^{-1}$ (UVW2), that correspond to $16.47\pm0.03$ mag, $16.94\pm0.05$ mag and $16.63\pm0.03$ mag, respectively.

\section{Results} \label{par:results}
Figure \ref{fig:BAT_radio_SMARTS} shows, with the same time scale, the X-ray and radio light curves of Swift J1753.5-0127. The time interval in which we also had OIR
observations is marked with a horizontal line. The peculiar
morphology of the X-ray light curve has already been discussed in \S \ref{par_1753}. VLA and MERLIN sampled intensively the
first part of the outburst, in which both the radio and the X-ray flux reached their peak. VLA also observed the source in a more recent phase of the outburst, in 2009
June. The WSRT and SMARTS observations instead focused on part of the slow-rise phase in 2007 July-September.\\
From an inspection of Figure \ref{fig:BAT_radio_SMARTS} we note that the morphology of the radio lightcurve is reminiscent of the X-ray one: the two curves
feature similar shapes.
Table \ref{tab:log_obs_radio_ALL} (in the appendix) reports the spectral indices $\alpha$ of the radio spectra for all the observations with detections at different
wavelengths (or with one detection and one upper limit). Focusing on the observations with detections at three (or more) frequencies (since they should give
the most reliable estimates of $\alpha$), we notice that five of them have an optically thin spectrum, one of them has an inverted spectrum while the remaining three are
consistent with having either optically thin or slightly-inverted spectra. All these observations have been performed within the first two months from the beginning of the
outburst. We do not observe any clear evolution of $\alpha$.
\begin{figure}
\begin{tabular}{c}
\resizebox{8.5cm}{!}{\includegraphics{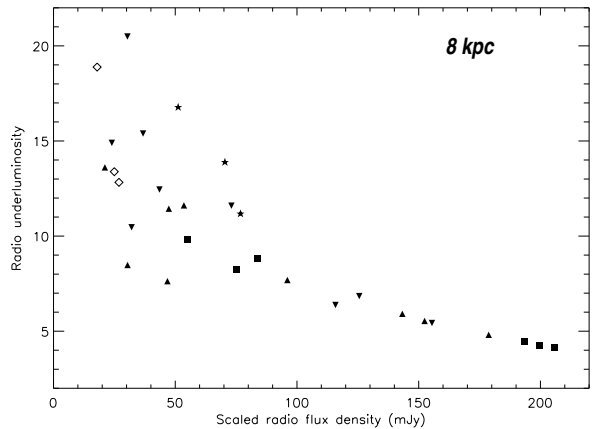}}
\end{tabular}
\caption{Radio underluminosity of Swift J1753.5-0127 as a function of the scaled radio flux density. We considered only the radio detections and not the
$3 \sigma$ upper limits. The radio underluminosity has been defined in \S \ref{par:gallo_rel}. The symbols mark different telescopes and frequencies,
as in Figure \ref{fig:BAT_radio_SMARTS}.}
\label{fig:radio_underlum}
\end{figure}
We will now focus on the radio and OIR observations that are simultaneous to Swift/XRT and/or RXTE observations (Figure \ref{fig:BAT_licu}).
The aim is to understand whether the jet is constantly less luminous than expected from the empirical correlations of
Gallo et al. (2006) and Russell et al. (2006) introduced in \S \ref{par:intro}.
\begin{figure}
\begin{tabular}{c}
\resizebox{8.5cm}{!}{\includegraphics{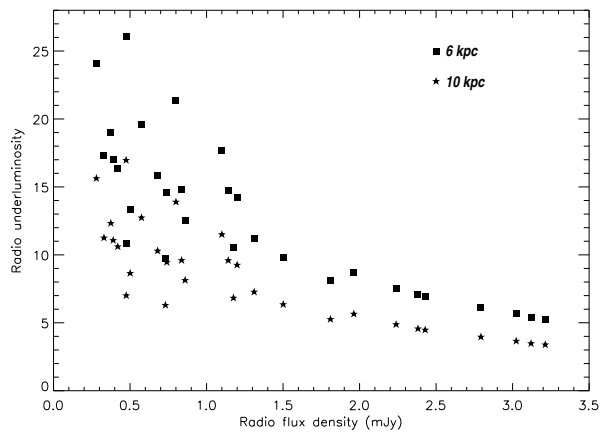}}
\end{tabular}
\caption{Radio underluminosity of Swift J1753.5-0127 as a function of the radio flux density density for two other possible distances to the source. Different symbols
mark the two distances (6 and 10 kpc). On the X axis we plotted the radio flux density and not the scaled radio flux density, as we did in Figure \ref{fig:Gallo_relation}
and Figure \ref{fig:radio_underlum}. We considered only the radio detections and not the $3 \sigma$ upper limits.}
\label{fig:underlum_dist}
\end{figure}

\subsection{Radio/X-ray correlation} \label{par:gallo_rel}
Cadolle Bel et al. (2007) tested whether the scaled radio flux of Swift J1753.5-0127 is consistent with that predicted by the empirical 
radio/X-ray correlation found for several sources, for a set of possible distances (1-15 kpc). The system was observed near the peak of the outburst on 2005 August 08,
when the X-ray flux was $F_{2-11 keV} = 1.5 \times 10^{-9}$ erg cm$^{-2} s^{-1}$. They concluded that, no matter which distance is assumed, the source
is less luminous in radio than expected from the correlation (at the same X-ray luminosity).
Figure \ref{fig:BAT_licu} shows, all along the outburst, when we had simultaneous or quasi-simultaneous (within 24 hours) X-ray and radio observations. The corresponding
dates are listed in Tables \ref{tab:log_obs_X} and \ref{tab:log_obs_radio}.\\ 
Figure \ref{fig:Gallo_relation} shows the BHCs sample published in Gallo et al. 2006 (the dashed line is the fit to their data 
$S_{radio} = 225 \cdot (S_{X})^{0.58\pm0.16}$), plus the outliers presented in Gallo (2007) and the data recently published
by McClintock et al. (2009) and Jonker et al. (2010) for the BHC H~1743-322. Our Swift J1753.5-0127 data points are also plotted. All the radio and X-ray fluxes have been
scaled to a distance of 1 kpc. 
Our campaign was performed over $\sim 4$ years, covering a broad range in X-ray flux (a factor $\sim 12$ between the peak and the minimum). Nevertheless, the lack of
coordinated observations at low X-ray fluxes (below a scaled flux of $\sim 1$ Crab) does not allow to obtain a statistically good fit to the data. For this reason,
we can only give a range of slopes $b$ and normalisations $k$ (for a relation of the form $S_{radio} = k (S_{X})^{b}$): $b \sim 1.0-1.4$ and $k \sim 6-13$ (see
the dashed-dotted and the dotted line in Figure \ref{fig:Gallo_relation}).
At a given X-ray luminosity, Swift J1753.5-0127 is less luminous than expected from the radio/X-ray correlation
by a factor $\sim 4-20$ (considering only the detections and not the upper limits). Figure \ref{fig:radio_underlum} shows, for all the simultaneous X-ray and radio
detections (reported in Table \ref{tab:log_obs_radio}), the corresponding radio underluminosity as a function of the scaled radio flux density. We define the radio
underluminosity as the ratio between the scaled radio flux density expected from the correlation of Gallo et al. (2006) and the scaled radio flux density that we actually
measured. The radio underluminosity has been plotted for all the available telescopes and frequencies. Figure \ref{fig:underlum_dist} shows how the radio
underluminosity is affected by the distance to the source (see the discussion in Cadolle Bel et al. 2007 and Zurita et al. 2008) that we are using to scale
the radio and the X-ray fluxes. Considering two other possible distances to Swift J1753.5-0127 (6 and 10 kpc), the radio underluminosity is in the range $\sim 3-26$.

\subsection{OIR/X-ray correlation} \label{par:russell_rel}
\begin{figure*}
\begin{tabular}{c}
\resizebox{12cm}{!}{\includegraphics{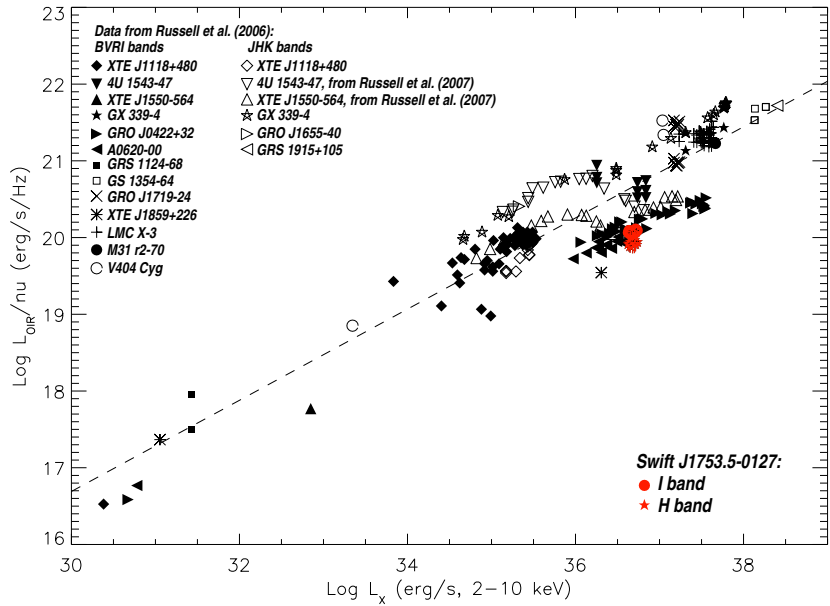}}
\end{tabular}
\caption{The BHCs sample of Russell et al. (2006), with the addition of the Swift J1753.5-0127 data. We also included data from the BHCs XTE J1550-564 and 4U 1543-47 (near
IR only), from Russell et al. (2007). The dashed line is the fit to the Russell et al. (2006) data, a relation of the
form $L_{OIR} \propto L_{X}^{0.61\pm0.02}$ in the ($L_X,L_{OIR}$) plane. Our data of Swift J1753.5-0127 are marked in red, for the two bands (I and H). We did not plot
the error bars for clarity.}
\label{fig:Russell_relation}
\end{figure*}

Figure \ref{fig:Russell_relation} shows the large sample of BHCs used by Russell et al. (2006) to obtain a correlation between the OIR and X-ray luminosities
in the LHS and in quiescence. We also included the data points from our simultaneous X-ray and OIR observations of Swift J1753.5-0127 and the data points from the BHCs
XTE J1550-564 and 4U 1543-47 (near IR only), from Russell et al. (2007).
At a given X-ray luminosity, Swift J1753.5-0127 is less luminous than expected from the OIR/X-ray correlation by a factor $\sim 3.5-4.1$ for
the I band and $\sim 4.9-5.7$ for the H band. In Figure \ref{fig:Russell_relation} the source is located with the other BHCs, although it is on the edge of the
cluster that they form. The SMARTS observations were performed in a short time interval ($\sim 2$ months) in which the X-ray and the I band fluxes
varied at most by a factor $\sim 1.3$ (Table \ref{tab:log_obs_X}) and $\sim1.2$ (Table \ref{tab:log_obs_smarts}), respectively. This tells that
defining an OIR underluminosity (as the ratio between the OIR luminosity expected from the correlation of Russell et al. (2006) and the OIR luminosity that we
actually measured) would not give any additional information.

\subsection{Spectral energy distributions} \label{par:SED}
\begin{figure*}
\begin{tabular}{c}
\resizebox{15cm}{!}{\includegraphics{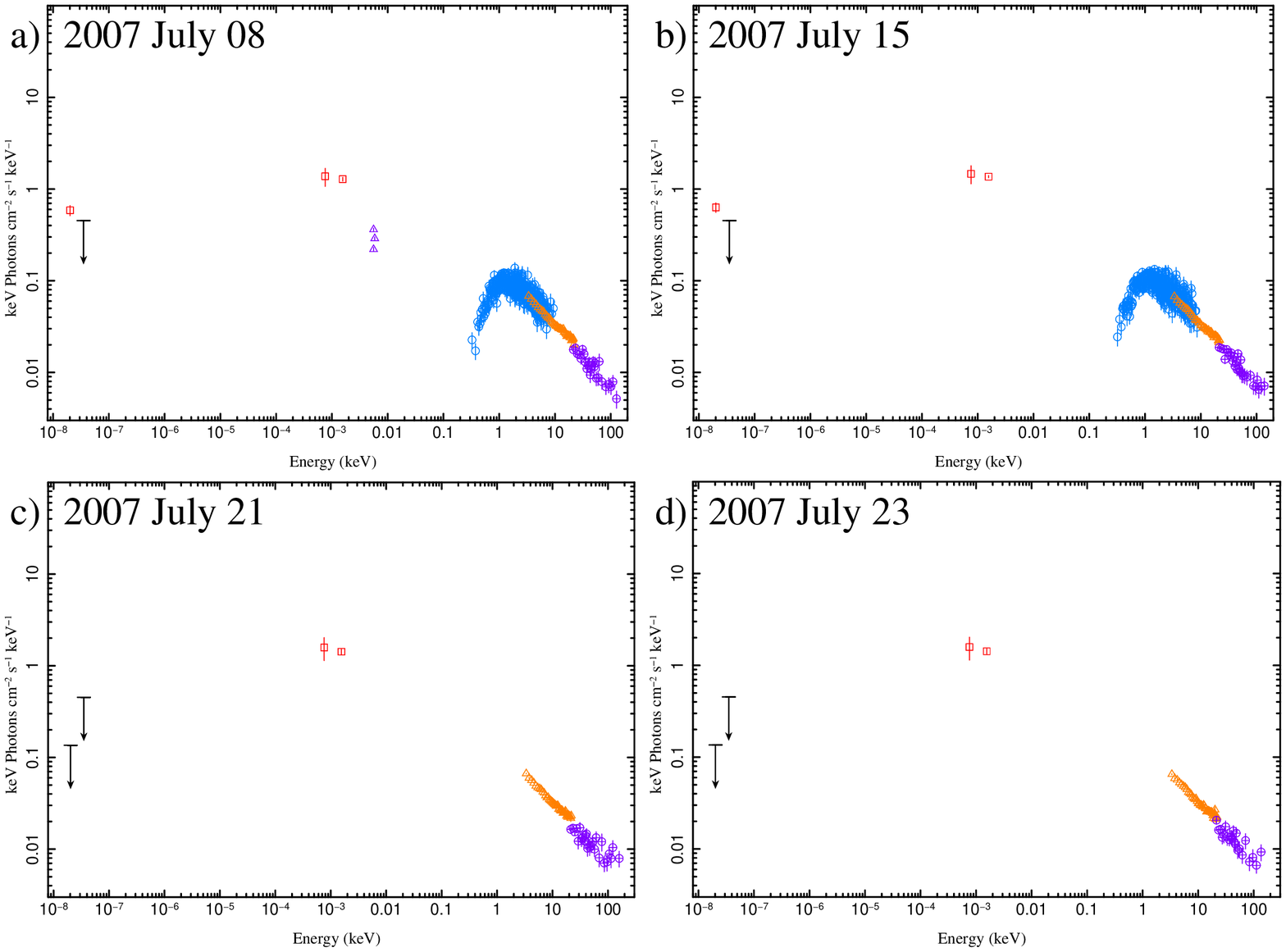}}
\end{tabular}
\caption{SEDs of Swift J1753.5-0127 on 2007 July 08, 15, 21 and 23 (panels a, b, c and d, respectively). We are showing the observed OIR, ultraviolet and X-ray fluxes,
without correcting for the absorption.
The red squares and the black arrows on the left-hand side of each panels mark the radio WSRT observations (detections at 4.9 GHz and $3 \sigma$
upper limit at 8.5 GHz). The red squares (at an energy of $\sim 10^{-3}$ keV) represents the H and I bands detections with SMARTS. The Swift/UVOT detections
(ultraviolet, available only for panel a) are marked with purple triangles. X-ray points are marked with blue open circles
(Swift/XRT, not available for panels c and d), orange open triangles (RXTE/PCA) and purple open circles (RXTE/HEXTE). See \S \ref{par:data} for the
details on the data reduction and Tables \ref{tab:log_obs_X}, \ref{tab:log_obs_radio} and \ref{tab:log_obs_smarts} for the values of the used fluxes. Panel c): we used
WSRT and RXTE data from 2007 July 22 and SMARTS data from 2007 July 21. Panel d): we used WSRT and RXTE data from 2007 July 22 and SMARTS data from 2007 July 23.}
\label{fig:SED_all}
\end{figure*}
We created 4 quasi-simultaneous (all observations were done within $\sim32$ hours) broadband SEDs. To get the de-reddened/un-absorbed spectrum (e.g. in the X-ray band)
we need to assume a continuum model. Since we are not modelling the SEDs, a conservative approach suggests to present the observed OIR, ultraviolet and X-ray fluxes,
without correcting for the absorption. Figure \ref{fig:SED_all} (panel a) shows the
SED obtained around 2007 July 08 which has the best spectral coverage, since it includes also ultraviolet data.
The radio--OIR spectral index $\alpha$ calculated for the SEDs in the panels a and b (using the radio flux density at 4.9 GHz and the de-reddened H and I-band flux
densities in Tables \ref{tab:log_obs_radio} and \ref{tab:log_obs_smarts}) is
slightly inverted, $\alpha = 0.12\pm0.01$ ($\sim 0.10$ without considering the I band). However, considering only the radio detection at 4.9 GHz and the
$3 \sigma$ upper limits at 8.5 GHz we find optically thin spectra with spectral indices $\alpha \lesssim -0.48$ and $\alpha \lesssim -0.61$ for the SEDs
in panels a and b, respectively. 
The SEDs from 2007 July 21 and 2007 July 23 are reported in Figure \ref{fig:SED_all} (panels c and d, respectively). They show the radio upper limits (at 4.9 and 8.5 GHz)
obtained on 2007 July 22, which do not allow us to calculate the spectral index $\alpha$ of the radio spectrum. The non detection at 4.9 GHz, compared to the
detection at the same frequency during the previous WSRT observation (on 2007 July 15), tells that the radio flux density lowered at least by a factor $\sim5$.

\section{Discussion} \label{par:discussion}
A primary aim of our observing campaign was to test whether Swift J1753.5-0127 features a faint jet in all the phases of its ongoing (at least at the
moment of writing this paper) outburst. Figure \ref{fig:Gallo_relation} firmly proves that the jet is {\it always} fainter than expected from the empirical radio/X-ray
correlation (Gallo et al. 2006).\\
It is worth to mention that the position of the bright outliers in Figure \ref{fig:Gallo_relation} (with X-ray scaled flux above $\sim 1$ Crab), because of the
partially-quenched radio emission, is reminiscent of the position of some BHCs in the HSS (see Gallo et al. 2003). Nevertheless, all the outliers
in Figure \ref{fig:Gallo_relation} (including Swift J1753.5-0127) have been observed in the LHS (Gallo 2007; Zhang et al. 2007, Jonker et al. 2010).

In \S \ref{par:gallo_rel} we defined a parameter, the ``radio underluminosity''. An inspection of Figure \ref{fig:radio_underlum} shows that there is no correlation between
the observing frequency and the radio underluminosity and that this parameter decreases for an increasing radio flux. This trend is expected, since the slope of the best-fit
of the Swift J1753.5-0127 data points is steeper than the slope reported in Gallo et al. (2006). The BHC H~1743-322 has been found to follow an opposite behaviour
(its radio underluminosity increases for an increasing radio flux), since the slope of the best-fit correlation reported in Jonker et al. (2010) is $b_{J09}=0.18\pm0.01$.\\
To test whether the radio underluminosity in Swift J1753.5-0127 is not just an artifact due to our assumption on the distance to the source ($D = 8$ kpc), in Figure
\ref{fig:underlum_dist} we plotted the radio underluminosity for two other possible distances (6 and 10 kpc). This parameter varies in the range $\sim 3-26$: the radio
emission from Swift J1753.5-0127 is {\it always} less luminous than expected from the radio/X-ray correlation, no matter which distance to the source we assume
(in the range $6-10$ kpc).

In Figure \ref{fig:Russell_relation} we tested for the first time whether the OIR emission from Swift J1753.5-0127 is less luminous than expected from the correlation of
Russell et al. (2006). Our coordinated X-ray/OIR observations spanned $\sim2$ months. We found that the OIR emission is less luminous then expected by a factor
$\sim 3.5-4.1$ in the I band and $\sim 4.9-5.7$ in the H band.
Swift J1753.5-0127 lies in Figure \ref{fig:Russell_relation} at the edge of the cluster formed by the other BHCs.

Coriat et al. (2009) recently reported the results from coordinated OIR and X-ray observations of four outbursts of the BHC GX 339-4. They discovered that even in
the HSS the X-ray and the OIR fluxes are correlated, with a slope similar to the LHS but with a lower normalisation. They also suggested a disc origin for the OIR emission
of GX 339-4 in the HSS. Interestingly, the position of Swift J1753.5-0127 in Figure \ref{fig:Gallo_relation} is reminiscent of the position of
the HSS points of GX 339-4 in Figure 3 of Coriat et al. (2009). However, the same origin can not be invoked to explain the outliers of the radio/X-ray correlation, since the
radio emission is not coming from the accretion disc.

The majority of the spectral indices in Table \ref{tab:log_obs_radio} (see Table \ref{tab:log_obs_radio_ALL} for the complete list) are consistent with an optically thin radio spectrum, also the ones obtained from the WSRT observations on 2007 July 08 and 15 (however their determination is highly uncertain since only one detection and a $3 \sigma$ upper limit are available).
Flat or slightly-inverted spectra are common for BHCs in the LHS (Fender 2001;
Fender et al. 2001) and are considered the signature of the compact-jet spectrum extending from radio to higher frequencies (Markoff et al. 2003). This suggests that, at least in those observations in which the radio spectrum is optically thin, we are observing blobs of plasma downstream in the jet, rather than the compact-steady jet itself (see Fender et al. 2004 and Jonker et al. 2010). As a consequence, the radio underluminosity of the compact jet is higher than the one that we actually measured, at least in the observations in which the radio spectrum is optically thin.
We could precisely determine $\alpha$ (with radio detections at three or more frequencies) only within $\sim 2$ months from the outburst peak and in
all the cases (but one, on 2005-07-26) the value of the spectral index is consistent with being negative. Fender et al. (2004) and Casella et al. (2005) suggested that these blobs (characterized by negative radio spectral indices) are ejected during the transition from the hard intermediate state to the soft intermediate state. However, optically thin radio spectra have already been seen during LHS-only outbursts (e.g. in GS 1354-64, Brocksopp et al. 2001; XTE J1118+480, Brocksopp et al. 2010; GRO J1719-24 and possibly in V404 Cyg, Brocksopp et al. 2004) or in the LHS of ``normal'' BHC outbursts (e.g. in H~1743-322, Jonker et al. 2010). This shows that it is indeed possible that
Swift J1753.5-0127 had a major ejection near the peak of the outburst, even without leaving the LHS. If so, the optically thin radio emission would probably come from the superposition of a faint compact-steady jet and the discrete ejecta (see Jonker et al. 2010 and Brocksopp et al. 2010 for a discussion).\\
The radio/X-ray correlation of Gallo et al. (2003) applies to compact-steady jets in the LHS. Thus, any derivation of the slope of the radio/X-ray correlation based on data points characterized by optically thin spectra can not be directly compared.
However, in this paper we presented the correlation using radio data in which the contribution of the discrete ejecta probably dominates. Considering the six spectral indices in Table \ref{tab:log_obs_radio} obtained with detections at three frequencies (or more), two of them are consistent with being flat or slightly inverted. This suggests that a compact-steady jet is indeed on, despite the optically thin emission from the discrete ejecta sometimes dominates. Figure \ref{fig:Gallo_factor17} shows a zoom of Figure
\ref{fig:Gallo_relation}, in which only the data points from Swift J1753.5-0127 are reported. The data points from observations characterized  
by optically thin spectra are located in the diagram together with the data points from observations with $\alpha \ge 0$.
For these reasons, we presented the radio/X-ray correlation using all the available coordinated observations, bearing in mind that our determination of $b$ should be treated with care, at least when the radio spectrum is optically thin.\\
The negative spectral indices in Table \ref{tab:log_obs_radio}
also suggest that the OIR emission does not originate in the jet (so the radio and the OIR come from unconnected regions), but for example is produced by the reprocessing of the X-rays in the outer regions of the accretion disc or from the outer regions of the accretion disc itself. The contribution of the companion star is most likely negligible in the OIR bands (Cadolle Bel et al. 2007, Zurita et al. 2008, Hynes et al. 2009), since the system probably hosts a late type K or M star (Cadolle Bel et al. 2007), more specifically an M2V-type star (Zurita et al. 2008).\\ 
The slope of the OIR/X-ray correlation can be used to test whether the the reprocessing or the emission from the outer disc are viable mechanisms
(van Paradijs \& McClintock 1994, but see Coriat et al. 2009). Unfortunately during our OIR campaign the X-ray and OIR fluxes did not considerably
vary (see \S \ref{par:russell_rel}), so we can not fit any correlation to the data.\\

\subsection{The scatter of the radio/X-ray correlation}
\begin{figure}
\begin{tabular}{c}
\resizebox{8.6cm}{!}{\includegraphics{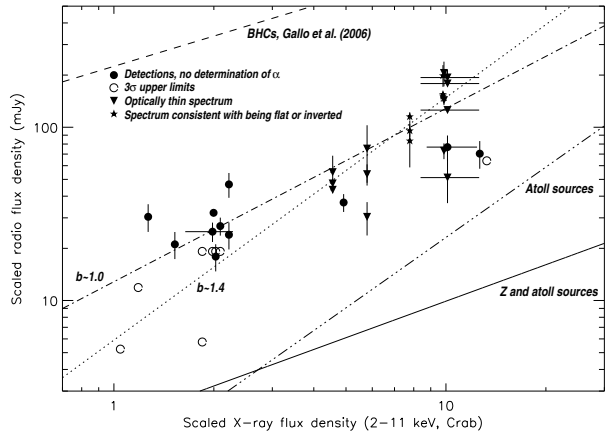}}
\end{tabular}
\caption{Zoom of Figure \ref{fig:Gallo_relation}, in which only the data points from Swift J1753.5-0127 are reported. A key to the different symbols is also shown. We determined the radio spectral index only for those observations with detections at three (or more) frequencies.
The dashed, dotted-dashed and dotted lines have already been defined in \S \ref{par:gallo_rel} and in the caption of Figure \ref{fig:Gallo_relation}. 
The solid line and the dashed-dotted-dotted line represent the best fits reported in Figure 2 of Migliari \& Fender (2006) for the atoll and Z sources ($k \sim 1.98$,
$b \sim 0.7$) and the atoll sources only ($k \sim 1.03$, $b \sim 1.35$), respectively.}
\label{fig:Gallo_factor17}
\end{figure}

From the fitting of our data we could only give a range of values in which the slope and the normalisation of the radio/X-ray correlation vary. Further observations
when Swift J1753.5-0127 is fading to quiescence and in the quiescent state will be fundamental to better constrain these parameters.\\
A precise determination of the slope $b$ of the correlation is particularly important. In fact, assuming a certain scaling between the radio power and the jet power
(for optically thick jets, following Blandford \& K\"onigl 1979, $L_R \propto L_J^{1.4}$) and a correlation between the accretion-powered X-ray luminosity $L_X$ and the
radio luminosity $L_R$, we can estimate the radiative efficiency of the jet (see Fender, Gallo \& Jonker 2003 and Migliari \& Fender 2006). In other words,
one can infer how the power carried by the jet $L_J$ scales with the mass accretion rate $\dot{m}$. 
Figure \ref{fig:Gallo_factor17} shows a zoom of Figure \ref{fig:Gallo_relation}, in which only the data points from Swift J1753.5-0127 are reported. It  also shows
the best fit correlations, from Figure 2 in Migliari \& Fender (2006), for a sample of atoll sources and atoll and Z sources (NSXBs accreting respectively at {\it low}
and {\it high} Eddington rate with X-ray luminosities $L_{X,atoll} \lesssim 0.5 L_{Edd}$ and $L_{X,Z} \gtrsim 0.5 L_{Edd}$, see van der Klis 2006 for a review).
We note that the upper range of possible slopes for the radio/X-ray correlation in Swift J1753.5-0127 is comparable to that measured for the neutron star system
(but see Tudose et al. 2009 for the atoll source Aql X-1).

The possibility that the outliers to the radio/X-ray correlation could follow a relation characterized by a similar slope to the ``normal'' BHCs but with a different
normalisation (Corbel et al. 2004; Gallo 2007) has important implications. For instance, it implies that the slope of the radiative efficiency does not vary in different
sources but some other parameters and factors might interfere in regulating the energy output in the form of an outflow. K\"{o}rding et al. (2006a) presented a method to
measure the mass accretion rate $\dot{m}$ from the radio luminosity. The presence of sources characterized by similar accretion properties but very different outflows
would be a major problem for the applicability of their method.\\
Pe'er \& Casella (2009) presented a model for emission from jets in X-ray binaries, in which electrons are accelerated only once at the base of
the jet (at variance with other jet models, in which multiple accelerations occur; see e.g. Maitra et al. 2009, Jamil, Fender \& Kaiser 2010). In the model, a jet magnetic
field above a critical value $B_{cr} \approx 10^{5} G$ would cause a quenching of the jet, without influencing the energy output in the X-ray band (Casella \& Pe'er 2009).\\
In the case of BHCs, the dependence of the outflows on the black-hole spin should also be analysed. However, no systematic study on how the spin couples with the
phenomenology of the ejecta has been reported yet (we address the reader to Fender, Gallo \& Russell 2010, submitted). Models involving a role of the spin of the compact
object should be able to explain the fact that AGN seem to feature a similar inflow/outflow coupling to BHCs (when a mass dependence is considered; Merloni et al. 2003,
Falcke et al. 2004).\\
The scatter in the radio X-ray correlation could also be analysed in terms of de-boosting effects of jets with different values of the bulk Lorentz factor $\Gamma$ and different inclination angles, as already discussed in Gallo et al. (2003)
and Heinz \& Merloni (2004). In fact, if a jet were highly relativistic (with $\Gamma > 2$) and not pointing towards the observer, it would be de-boosted. Observationally, it would appear dimmer than it intrinsically is.\\
Compact-steady jets are thought to be mildly relativistic (with bulk Lorentz factor $\Gamma < 2$; Gallo et al. 2003, but see Casella et al. 2010) in the LHS below about $1\%$ of $L_{Edd}$ (e.g. Fender et al. 2003) while major relativistic ejections
($\Gamma \geq 2$) are in some cases associated with the transition from the hard to the soft states (Fender et al. 2004). These transitions occur at various $L_X$ but usually above a few per cent of $L_{Edd}$. This suggests that it is reasonable to expect the bulk Lorentz factor to vary when the source approaches the transition from the hard to the soft states, even within the LHS. In a subsequent paper (Soleri \& Fender, in prep.) we will discuss how a possible dependence of the Lorentz factor $\Gamma$ on the accretion-powered luminosity could qualitatively reproduce the observed scatter around the radio/X-ray correlation.

\subsection{Swift J1753.5-0127: a peculiar source}
As we already mentioned in \S \ref{par_1753}, no dynamical measurement of the mass of the compact object in Swift J1753.5-0127 is available.
Although Swift J1753.5-0127 is fainter in radio than ``standard'' BHCs by a factor $\sim 4-20$, other dynamically-confirmed BHCs can be as radio dim as it. For example
the BHC XTE J1550-564 contains a black hole with mass $9.6 \pm 1.2 M_{\odot}$ (Orosz et al. 2002) but it occupies a region of the diagram in
Figure \ref{fig:Gallo_relation} together with the other outliers. Similar conclusions can be drawn for the BHC XTE J1650-500, that contains a black hole with mass
probably in the range $4-7.3 M_{\odot}$ (Orosz et al. 2004). Figure \ref{fig:Gallo_factor17} shows that Swift J1753.5-0127 lies in a region of the ($S_X, S_{radio}$)
diagram in between the correlation found by Gallo et al. (2006) for BHCs and the correlations reported in Migliari \& Fender (2006) for NSXBs, so we can not
consider its position a hint for the nature of the compact object in the binary system.
The slope of the radio/X-ray correlation $b$ fitted for the Swift J1753.5-0127 data lies in the interval $\sim 1.0-1.4$. Considering this broad range and the
fact that both in NSXBs and BHCs the determination of $b$ is highly uncertain and (in some cases) source dependent (Migliari \& Fender 2006; Corbel et al. 2008,
Jonker et al. 2010), we can not use the value of $b$ as an estimator for the nature of the compact object in the system. 

\section{Conclusions} \label{par:conclusions}
We observed the BHC Swift J1753.5-0127 with a campaign of coordinated multiwavelength observations in the radio, OIR and X-ray bands.
Following the source for $\sim$4 years we could clearly confirm that it features a jet that is fainter than expected from the empirical correlation between the radio
and the X-ray scaled fluxes (Corbel et al. 2003; Gallo et al. 2003, 2006). We also verified that the Swift J1753.5-0127 is only slightly fainter in OIR than
expected from the correlation of Russell et al. (2006): in the ($L_X,L_{OIR}$) plane the source clusters with other BHCs.\\
From the analysis of the SEDs in Figure \ref{fig:SED_all} we inferred that probably the jet is not responsible for the OIR emission. Viable mechanisms are either
reprocessing of the X-rays in the outer regions of the accretion disc or emission from the accretion disc itself while the possibility that the OIR emission comes from
the companion star seems unlikely.\\
The optically thin radio spectra that we obtained suggest that the radio emission originates in blobs of plasma ejected near the peak of the outburst.
However, the presence of observations characterized by flat or slightly inverted spectra is a hint that a compact, steady jet is indeed on, although the radio emission is often dominated by the discrete ejecta detached from the core.

The best-fit parameters to the Swift J1753.5-0127 data in the ($S_X, S_{radio}$) plane could only be poorly constrained, because of the lack of observations at low  
($< 1$ Crab) scaled X-ray flux. One must also keep in mind that any determination of the slope of the radio/X-ray correlation from radio observations characterized by
optically thin spectra optically thin spectra should be treated with care. We discussed the importance of a precise determination of both the slope $b$ and the normalisation $k$ of the radio/X-ray correlation to infer the radiative efficiency of the jet. The possibility that (at least some of) the outliers of the correlation could be fitted
using a relation characterized by the same slope $b$ as the majority of the BHCs but a lower normalisation suggests that some parameters (e.g. the spin of the black hole and
the jet magnetic field) might play a role in regulating the disc/jet coupling and the energy output in the form of a jet.

We also briefly discussed on the nature of Swift J1753.5-0127, since the mass of the accretor  has not been measured. From our data we can not draw any conclusion on
the nature of the compact object. Although the possibility that the system is a NSXBs can not be excluded, the detection of a hard power-law tail in the
X-ray/$\gamma$-ray energy spectrum up to $\sim$600 keV still constitutes the strongest hint that Swift J1753.5-0127 is a BHC.

\section*{Acknowledgments}
PS acknowledges support from an NWO (Netherlands Foundation for Scientific Research) VIDI awarded to R. Fender and a NWO Spinoza awarded to
M. van der Klis. The authors thank the Principal Investigator of the Swift mission, Neil Gehrels, for approving and scheduling the Swift ToO (Target of
Opportunity). We thank Elena Gallo and David Russell for providing the original data for Figures \ref{fig:Gallo_relation} and Figure
\ref{fig:Russell_relation}. We also thank Simone Migliari for providing the correlations found for NSXBs which have been shown in Figure \ref{fig:Gallo_factor17}.
We are grateful to the anonymous referee for the useful comments and suggestions.
TB acknowledges support from the Italian Space Agency through grant I/008/07/00 and from the International Space Science Institute (IHSSI).
JCAM-J is a Jansky Fellow of the National Radio Astronomy Observatory. The National Radio Astronomy Observatory is a facility of the
National Science Foundation operated under cooperative agreement by Associated Universities, Inc.
The WSRT is operated by the ASTRON (Netherlands Institute for Radio Astronomy) with support from the NWO.
MERLIN, a UK National Facility, is operated by the University of Manchester at Jodrell Bank Observatory on behalf of STFC (Science and Technology Facilities Council). 
This research has made use of data obtained through the High Energy Astrophysics Science Archive Research Center Online Service, provided by the NASA (National
Aeronautics and Space Administration) Goddard space flight center.



\appendix
\section[]{Radio and OIR observations} \label{app_tables}
Here we report three tables in which we list all the radio and OIR observations reported in Figure \ref{fig:BAT_radio_SMARTS} (upper panel).
\begin{table*}
\centering
\caption{Log of all the VLA (proposal codes AR570, AR572, AR603, AM986, AT320 and S7810), MERLIN and WSRT observations used to plot the radio lightcurve in
Figure \ref{fig:BAT_radio_SMARTS}.}
\label{tab:log_obs_radio_ALL}
\begin{tabular}{c c c c c c c c}
\hline
\multicolumn{1}{c}{{\bf Radio observations}} & \multicolumn{6}{c}{Flux densities (mJy)}                                           & Spectral index $\alpha$ \\
           &     MERLIN          &                    \multicolumn{3}{c}{VLA}                   &       \multicolumn{2}{c}{WSRT}  &                         \\
Date       &    1.7 GHz          &     1.4 GHz   &     4.8 GHz              &      8.4 GHz      &   4.9 GHz           &   8.5 GHz &                         \\ 
\hline
2005-07-03 & $2.2\pm0.2$ 	 &	-	 &	-		    &	    -	        &      - 	      &      -	  &           -             \\
2005-07-04 & $1.1\pm0.2$	 &	-	 &	-		    &	    -	        &      - 	      &      -	  &           -             \\ 
2005-07-06 & $1.2\pm0.2$	 &	-	 &	-		    &	    -	        &      - 	      &      -	  &	      -             \\
2005-07-07 & $1.0^{a}$           &	-	 &	-		    &	    -	        &      - 	      &      -	  & 	      - 	    \\
2005-07-08 & $0.8\pm0.3$	 & $3.02\pm0.33$ & $2.79\pm0.05$            & $1.96\pm0.04$     &      - 	      &      -	  &	$-0.30\pm0.27$      \\
2005-07-10 &	-		 & $3.21\pm0.52$ & $2.24\pm0.14$	    & $1.14\pm0.12$     &      - 	      &      -	  &	$-0.59\pm0.30$      \\
2005-07-15 &    -                & $3.08\pm0.46$ & $2.76\pm0.14$            & $2.65\pm0.26$     &      - 	      &      -	  &	$-0.084\pm0.007$    \\
2005-07-19 &	-		 & $3.12\pm0.45$ & $2.38\pm0.13$	    & $2.42\pm0.05$     &      - 	      &      -	  &	$-0.12\pm0.37$      \\
2005-07-26 &	-		 & $1.31\pm0.39$ & $1.50\pm0.13$	    & $1.81\pm0.09$     &      - 	      &      -	  &	$0.26\pm0.09$       \\
2005-08-03 &	-		 & $1.18\pm0.43$ & $0.84\pm0.12$	    & $0.47\pm0.10$	&      - 	      &      -	  &     $-0.42\pm0.31$      \\
2005-08-07 &	-		 &	 -	 &	-		    & $0.57\pm0.07$     &      - 	      &      -	  &	      -             \\
2005-08-11 &    -                & $0.86\pm0.21$ & $0.74\pm0.05$	    & $0.68\pm0.01$	&      -	      &      -	  &     $-0.14\pm0.01$      \\
2005-08-12 &       -             &	  - 	 &      -                   & $0.97\pm0.21$     &      -              &      - 	  &           -             \\
2005-08-16 &       -             &        -	 &      -                   & $0.56\pm0.08$     &      -              &      - 	  &           -             \\
2005-08-18 &       -             &        -	 &      -                   & $0.65\pm0.05$     &      -              &      - 	  &           -             \\
2005-08-20 &       -             &        -	 &      -                   & $0.55\pm0.07$     &      -              &      - 	  &           -             \\
2005-08-24 &       -             & $0.66\pm0.45$ &  $0.57\pm0.09$           & $0.62\pm0.07$     &      -              &      - 	  &     $0.09\pm0.10$       \\
2005-08-29 &       -             & $1.05\pm0.32$ &  $0.52\pm0.07$           & $0.64\pm0.07$     &      -              &      - 	  &     $-0.005\pm0.435$    \\
2005-09-04 &       -             &       -	 &      -                   & $0.39\pm0.10$     &      -              &      - 	  &           -             \\
2005-09-11 &	   -		 &	 -	 &   $0.73\pm0.12$	    & $0.37\pm0.07$     &      -	      &      -    &    $-1.20^{b}$          \\
2005-09-18 &       -             &	 -	 &   $0.41\pm0.07$          & $0.45\pm0.09$     &      -	      &      -    &     $0.17^{b}$          \\
2005-09-23 &       -             &	 -	 &   $0.45\pm0.06$          &       -           &      -              &      - 	  &           -             \\
2005-10-07 &       -             &	 -	 &   $0.45\pm0.06$          &       -           &      -              &      - 	  &           -             \\
2005-10-17 &       -             &	 -	 &   $0.24^{a}$             &       -           &      -              &      - 	  &           -             \\
2005-10-22 &	   -		 &	 -	 &  $0.33\pm0.06$	    &       -  		&      -	      &      -    &	      -	            \\
2005-10-24 &       -             &	 -	 &  $0.36\pm0.06$           &       -           &      -	      &      -    &	      -	            \\
2005-11-19 &	   -		 &	 -	 &  $0.48\pm0.09$	    &       -  		&      -	      &      -    &	      -  	    \\
2005-12-23 &       -       	 &	 -	 &  $0.21\pm0.05$           &       -   	&      -	      &      -    &	      -             \\
2005-12-31 &       -      	 &	 -	 &  $0.27\pm0.05$           &       -   	&      -	      &      -    &	      -             \\
2006-01-08 &       -     	 &	 -	 &  $0.31\pm0.06$           &       -    	&      -	      &      -    &	      -             \\   
2006-01-14 &       -     	 &	 -       &  $0.35\pm0.06$           & $0.23\pm0.06$	&      -	      &      -    & $-0.75^{b}$             \\
2006-02-03 &       -    	 &	 -	 &  $0.16^{a}$              &       -   	&      -	      &      -    &	      -             \\
2006-02-06 &       -    	 &	 -	 &  $0.17^{a}$              &       -    	&      -	      &      -    &	      -             \\
2006-02-19 &       -    	 &	 -	 &  $0.07\pm0.05$           &       -    	&      -	      &      -    &	      -             \\
2006-03-11 &	   -		 &	 -	 &  $0.08^{a}$              &       -  		&      -	      &      -    &	      - 	    \\
2006-03-19 &	   -		 & $0.49\pm0.14$ &  $0.04\pm0.03$           &       -  		&      -	      &      -    & $-2.03^{b}$             \\
2006-04-12 &       -		 &	 -	 &  $0.18\pm0.06$           &       -   	&      -	      &      -    &	      -             \\
2006-04-29 &       -		 &	 -       &  $0.23^{a}$              &       -    	&      -	      &      -    &	      -             \\
2006-05-03 &       - 		 &	 -       &  $0.11\pm0.03$           &       -     	&      -	      &      -    &	      -             \\
2006-06-08 &       - 		 &	 -       &  $0.26\pm0.07$           &       -    	&      -	      &      -    &	      -             \\
2006-06-26 &       - 		 &	 -       &  $0.29\pm0.05$           &       -     	&      -	      &      -    &	      -             \\
2006-07-13 &       - 		 &	 -       &  $0.31\pm0.06$           &       -     	&      -	      &      -    &	      -             \\
2006-08-03 &	   -		 &	 -	 &  $0.19^{a}$              &       -  		&      -	      &      -    &	      - 	    \\
2006-08-04 &       -    	 &	 -	 &  $0.13^{a}$              &       -  		&      -	      &      -    &	      -             \\
2006-08-05 &       -    	 &	 -	 &  $0.22\pm0.04$           &       -   	&      -	      &      -    &	      -             \\
2006-08-09 &       -    	 &	 -	 &  $0.19^{a}$              &       -   	&      -	      &      -    &	      -             \\
2006-08-18 &       -     	 &	 -	 &  $0.19\pm0.05$           &       -  		&      -	      &      -    &	      -             \\
2006-08-28 &       -     	 &	 -	 &  $0.10\pm0.05$           &       -  		&      -	      &      -    &	      -             \\
2006-09-09 &       -    	 &	 -	 &  $0.17\pm0.06$           &       -  		&      -	      &      -    &	      -             \\
2006-09-29 &       -    	 &	 -       &  $0.32\pm0.05$           &       -  		&      -	      &      -    &	      -             \\
2007-01-28 &       - 	         &	 -	 &  $0.24\pm0.05$           &       -  		&      -	      &      -    &	      -             \\
2007-02-08 &       - 	         &	 -	 &  $0.19\pm0.04$           &       -  		&      -	      &      -    &	      -             \\
2007-02-10 &       -         	 &	 -	 &  $0.35\pm0.04$           &       -  		&      -	      &      -    &	      -             \\
2007-07-01 &	   -		 &	 -	 &	 -		    &       -  		& $0.28\pm0.05$       & $0.3^{a}$ &   $\lesssim 0.13$       \\
2007-07-08 &	   -		 &	 -       &	 -		    &       -  		& $0.39\pm0.05$       & $0.3^{a}$ &   $\lesssim -0.48$      \\
2007-07-15 &	   -		 &	 -	 &	 -		    &       -  		& $0.42\pm0.05$       & $0.3^{a}$ &   $\lesssim -0.61$      \\
2007-07-22 &	   -	         &	 -       &	 -		    &       -  		& $0.09^{a}$          & $0.3^{a}$ &         -               \\
2009-06-09 &       -             &	 -       &        -                 & $0.50\pm0.02$	&      -	      &      -    &	    -               \\
2009-06-10 &       -             &       -       &        -                 & $0.42\pm0.05$	&      -	      &      -    &	    -               \\
2009-06-15 &       -             &       -       &        -                 & $0.40\pm0.02$	&      -	      &      -    &	    -               \\
\hline
\end{tabular}
\\
$^{a}$ $3 \sigma$ upper limit
;
$^{b}$ the errors on the slope could not be obtained, since we fitted two data point using a model with two free parameters
\end{table*}
\begin{table}
\centering
\caption{Log of all the SMARTS observations. The period during which they have been performed is marked with a horizontal line in the upper panel of
Figure \ref{fig:BAT_radio_SMARTS}. The observed flux densities have been de-reddened assuming $R_{v} = 3.1$ and $A_{v} = 3.1 \times E_{B-V} = 1.05$ ($E_{B-V} \sim 0.34$,
Cadolle Bel et al. 2007).}
\label{tab:OIR_all}
\begin{tabular}{c c c}
\hline
\hline
\multicolumn{3}{c}{{\bf SMARTS OIR observations}}         \\
           & \multicolumn{2}{c}{Flux densities (mJy)}     \\
           &   I band            &    H band              \\
\hline
2007-07-09 & $1.51\pm0.01$       &  $1.09\pm0.02$         \\
2007-07-14 & $1.57\pm0.06$       &  $1.12\pm0.02$         \\
2007-07-15 & $1.61\pm0.04$       &  $1.16\pm0.03$         \\
2007-07-17 & $1.46\pm0.03$       &  $1.10\pm0.02$         \\
2007-07-18 & $1.58\pm0.01$       &  $1.19\pm0.02$         \\
2007-07-21 & $1.46\pm0.09$       &  $1.10\pm0.03$         \\
2007-07-23 & $1.69\pm0.01$       &  $1.25\pm0.04$         \\
2007-07-24 &       -             &  $1.13\pm0.04$         \\
2007-07-26 & $1.64\pm0.03$       &  $1.24\pm0.03$         \\
2007-07-27 & $1.56\pm0.07$       &  $1.11\pm0.02$         \\
2007-07-28 & $1.66\pm0.02$       &  $1.15\pm0.02$         \\
2007-07-30 & $1.71\pm0.09$       &  $1.19\pm0.03$         \\
2007-07-31 & $1.51\pm0.07$       &  $1.12\pm0.02$         \\
2007-08-02 & $1.56\pm0.06$       &  $1.11\pm0.02$         \\
2007-08-03 & $1.54\pm0.06$       &  $1.09\pm0.02$         \\
2007-08-04 & $1.58\pm0.07$       &  $1.17\pm0.03$         \\
2007-08-05 & $1.54\pm0.01$       &  $1.23\pm0.05$         \\
2007-08-07 & $1.58\pm0.04$       &  $1.15\pm0.03$         \\
2007-08-11 & $1.43\pm0.02$       &        -               \\
2007-08-13 & $1.47\pm0.03$       &  $1.02\pm0.02$         \\
2007-08-14 & $1.56\pm0.01$       &  $1.16\pm0.03$         \\
2007-08-15 & $1.61\pm0.01$       &  $1.24\pm0.03$         \\
2007-08-16 & $1.63\pm0.02$       &  $1.16\pm0.02$         \\
2007-08-17 & $1.63\pm0.05$       &  $1.26\pm0.04$         \\
2007-08-18 & $1.64\pm0.02$       &  $1.19\pm0.03$         \\
2007-08-19 & $1.53\pm0.04$       &  $1.05\pm0.02$         \\
2007-08-21 & $1.56\pm0.04$       &  $1.29\pm0.03$         \\
2007-08-22 & $1.50\pm0.03$       &  $1.28\pm0.04$         \\
2007-08-23 & $1.71\pm0.03$       &  $1.28\pm0.03$         \\
2007-08-24 & $1.61\pm0.01$       &        -               \\
2007-08-29 & $1.63\pm0.05$       &  $1.11\pm0.02$         \\
2007-09-03 & $1.54\pm0.03$       &  $1.10\pm0.02$         \\
2007-09-04 & $1.44\pm0.08$       &  $1.12\pm0.04$         \\
2007-09-06 & $1.44\pm0.01$       &  $1.11\pm0.02$         \\
2007-09-07 & $1.56\pm0.07$       &  $1.22\pm0.03$         \\
2007-09-08 & $1.58\pm0.07$       &        -               \\
2007-09-09 & $1.51\pm0.03$       &  $1.09\pm0.02$         \\
2007-09-10 & $1.50\pm0.10$       &  $1.10\pm0.02$         \\
2007-09-12 & $1.47\pm0.05$       &  $1.11\pm0.03$         \\
2007-09-14 & $1.57\pm0.03$       &  $1.21\pm0.03$         \\
2007-09-19 & $1.60\pm0.06$       &  $1.13\pm0.02$         \\
2007-09-20 & $1.58\pm0.04$       &  $1.12\pm0.02$         \\
2007-09-21 & $1.56\pm0.04$       &  $1.05\pm0.02$         \\
2007-09-22 & $1.60\pm0.03$       &  $1.06\pm0.02$         \\
2007-09-24 & $1.63\pm0.02$       &  $1.17\pm0.02$         \\
2007-09-28 & $1.49\pm0.01$       &  $1.20\pm0.03$         \\
2007-09-29 & $1.61\pm0.01$       &  $1.21\pm0.03$         \\
2007-09-30 & $1.39\pm0.02$       &  $1.13\pm0.03$         \\
\hline
\hline
\end{tabular}
\end{table}
\label{lastpage}

\end{document}